%% file: sample-acmsmall.tex
\mathchardef\mhyphen="2D 

\documentclass[acmsmall]{acmart}

\usepackage[ruled]{algorithm2e}
\usepackage{float}
\usepackage{amsmath}
\usepackage{multirow}

\usepackage{array}
\newcolumntype{P}[1]{>{\centering\arraybackslash}p{#1}}

\newcommand{\todo}[1]{\textcolor{red}{\textbf{#1}}}
\newcommand{\adam}[1]{\textcolor{blue}{\textbf{#1}}}
\newcommand{\done}[1]{\textcolor{green}{\textbf{#1}}}
\newcommand{\lzy}[1]{\textcolor{black}{#1}}
\newcommand{\llzy}[1]{\textcolor{black}{#1}}
\newcommand{\lllzy}[1]{\textcolor{black}{#1}}
\newcommand{\lzyfinal}[1]{\textcolor{black}{#1}}

\AtBeginDocument{%
  \providecommand\BibTeX{{%
    \normalfont B\kern-0.5em{\scshape i\kern-0.25em b}\kern-0.8em\TeX}}}

\setcopyright{acmlicensed}
\acmJournal{TOIS}
\acmDOI{10.1145/3445029}

\begin{document} 

\title{Meta-evaluation of Conversational Search Evaluation Metrics}


\author{Zeyang Liu}
\email{zeyang.liu@nottingham.ac.uk}
\affiliation{%
  \institution{University of Nottingham}
  \country{UK}
}

\author{Ke Zhou}
\email{ke.zhou@nottingham.ac.uk}
\affiliation{%
  \institution{University of Nottingham \& Nokia Bell Labs}
  \country{UK}
}

\author{Max L. Wilson}
\email{max.wilson@nottingham.ac.uk}
\orcid{0000-0002-3515-6633}
\affiliation{%
  \institution{University of Nottingham}
  \country{UK}
}

\renewcommand{\shortauthors}{Liu et al.}

\begin{abstract}
Conversational search systems, such as Google assistant and Microsoft Cortana, enable users
to interact with search systems in multiple rounds through natural language dialogues. 
Evaluating such systems is very challenging given that any natural language responses could be generated, 
and users commonly interact for multiple semantically coherent rounds to accomplish a search task. 
Although prior studies proposed many evaluation metrics, the extent of how those measures effectively capture 
user preference remain to be investigated. In this paper, we systematically meta-evaluate a variety of conversational search metrics.
We specifically study three perspectives on those metrics: (1) \emph{reliability}: the ability to detect ``actual'' performance differences as opposed to those observed by chance; (2) \emph{fidelity}: the ability to agree with ultimate user preference; and (3) \emph{intuitiveness}: the ability to capture any property deemed important: adequacy, informativeness and fluency in the context of conversational search. 
By conducting experiments on two test collections, we find that the performance of different metrics vary significantly across different scenarios whereas consistent with prior studies, existing metrics only achieve weak correlation with ultimate user preference and satisfaction. METEOR is, comparatively speaking, the best existing single-turn metric considering all three perspectives.
We also demonstrate that adapted session-based evaluation metrics can be used to measure multi-turn conversational search, achieving moderate concordance with user satisfaction. 
To our knowledge, our work establish the most comprehensive meta-evaluation for conversational search to date.
   
\end{abstract}

\ccsdesc[500]{Information systems~Evaluation of retrieval results}
\ccsdesc[300]{Information systems~Retrieval effectiveness}
\ccsdesc[500]{Computing methodologies~Discourse, dialogue and pragmatics}

\keywords{conversational search, meta-evaluation, metric, discriminative power}

\maketitle

\section{introduction}
\input{content/1introduction.tex}

\section{related work}\label{sec:related work}
\input{content/2relatedwork.tex}

\section{metrics, data and systems} \label{sec:metric}
\input{content/3metrics.tex}

\section{Meta-evaluation in Single-turn Conversation}
\label{sec:metaevaluationsingleturn}
\input{content/4.0MetaevaluationSingleTurn}

\section{Meta-evaluation in Multiple-turn Conversation} \label{sec:multipleturn}

\input{content/5multiturnevaluation.tex}

\section{discussion} \label{sec:discussion}
\input{content/6discussion.tex}

\section{conclusion} \label{sec:conclusion}
\input{content/7conclusion.tex}
\bibliographystyle{ACM-Reference-Format}
\bibliography{sample-base}


\end{document}

%% file: content/1introduction.tex

\begin{figure}[t]
    \centering
    \includegraphics[width=15cm]{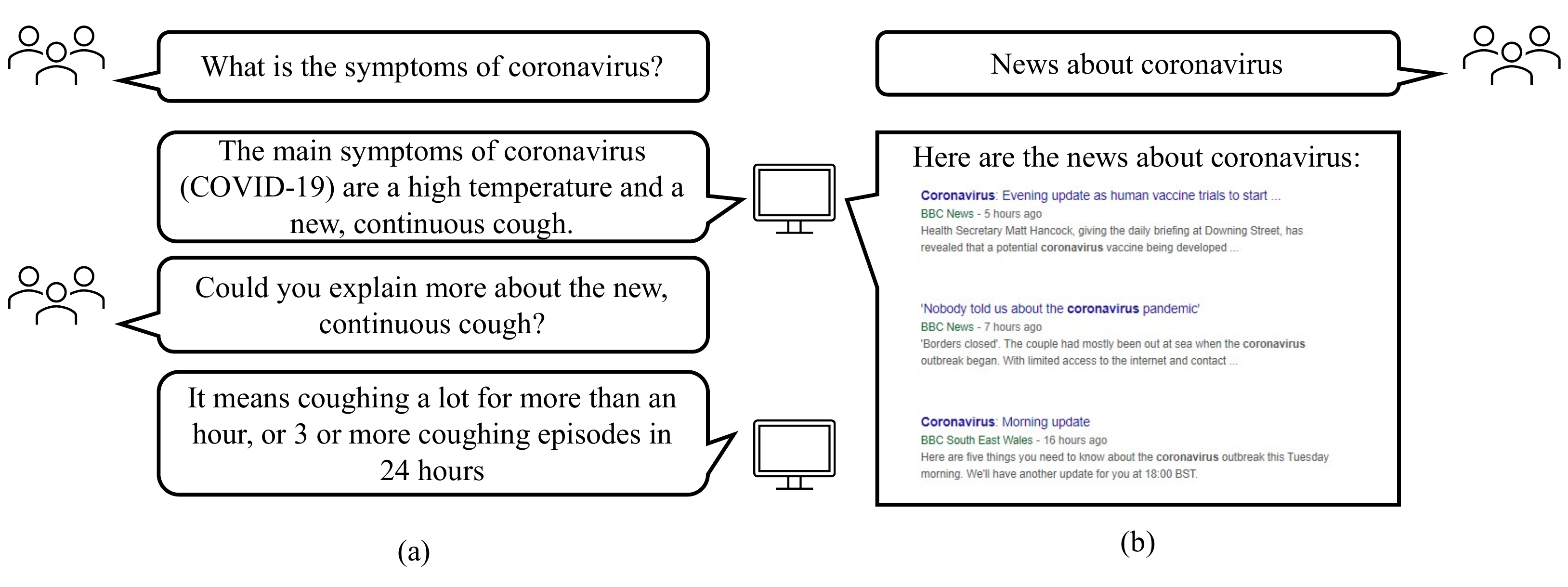}
    \caption{Two popular presentation strategies for conversational search: (a) Single Response in Single Turn (SRST); and (b) Multiple Responses in Single Turn (MRST).}
    \label{fig:examp_cv}
\end{figure}

The rapid growth of speech technology has significantly influenced the way that users interact with search systems to satisfy their information needs. \emph{Conversational search}, a new search paradigm where users interact with a search system using natural language conversations, has recently become very popular. Various voice assistants, such as Apple Siri, Google Assistant, and Microsoft Cortana, can be seen as typical applications of such a search paradigm. 
Unlike traditional search, these conversational search systems allow users to describe their search tasks in their natural language, and subsequently interact with the systems over multiple turns of dialogue exchanges.  It is worth noting, however, that the essence of a conversational search system is still an information retrieval (IR) system, which aims to provide the users with information that they want in order to satisfy their information needs \cite{radlinski2017theoretical}. 
%
Although these applications achieve remarkable success in the commercial market, the IR community still lacks a uniform approach to conversational systems. \citet{radlinski2017theoretical} define conversational search systems as `a system for retrieving information that permits a mixed-initiative back and forth between a user and agent'. 
Unlike traditional search engines that typically present a list of ten blue links, users have more flexible ways to express their information needs through performing dialogue-like interactions with the systems. For example, users may confirm their search target step by step over the multiple turns. Within each turn, there are two popular presentation strategies that were adopted by current conversational search systems, as shown in Figure~\ref{fig:examp_cv}.

\begin{itemize}
    \item \textbf{Single Response in Single Turn (SRST)}: in this scenario, the search agents provide one textual response to answer each of the users' questions. 
    This is typically available for hands-free user interactions.
    \item \textbf{Multiple Responses in Single Turn (MRST)}: in this scenario, the search system presents a list of textual answers like traditional information retrieval (IR) 
    systems while users can select and click one of these results from the list in each round. This is typically available when users have access to small mobile screens to further interact with the short and brief search results.  
\end{itemize}


Evaluation plays a pivotal role in designing and tuning search systems. However, due to the differences between traditional and conversational search, it is difficult to directly apply traditional search evaluation to measure conversational search systems.
There are several key differences, which make conversational search evaluation especially challenging:
\begin{itemize}
\item First, different from traditional search where document relevance can be assessed, conversational search systems can generate any responses (utterances) to address the information need. The large number of potential responses that can be generated makes relevance assessment of all responses infeasible. In addition, even if they are relevant to the information need, the generated responses may not share any words in common, or may even have different semantic meanings \cite{liu2016not}.
\item Second, as shown in Figure~\ref{fig:examp_cv}, the diversity of result presentation strategies increases the complexity of such evaluation. Given whether a single response or a ranked list of responses are returned, we need to design different metrics.
\item Last but not least, users tend to interact with conversational search systems with multiple turns (MT) of dialogue exchanges to address their information needs. It is essential to design metrics that can measure beyond single-turn (ST) conversations.  
\end{itemize}


Faced with such challenges, prior studies have proposed several automatic evaluation metrics, such as word-overlap based measures (e.g.,~BLEU \cite{papineni2002bleu}, METEOR \cite{banerjee2005meteor}) , word-embedding based metrics (e.g.,~Embedding Average \cite{serban2017hierarchical}, Soft Cosine Similarity \cite{sidorov2014soft} and BERTScore \cite{bert-score}), and \lllzy{learning-based metrics (e.g., BERT-RUBER \cite{ghazarian2019better})}, to measure system effectiveness. 
However, it is unclear which metrics are most suitable for different scenarios and whether they are trustworthy.
Similar to the focus of our paper, a few prior meta-evaluation studies \cite{liu2016not, novikova2017we, stent2005evaluating, kilickaya2016re} have compared these metrics. For example, \citet{liu2016not} evaluate the correlation between metrics and human annotations in the single-turn dialogue scenario. They indicate that both word-overlap based and word-embedding based metrics correlate weakly with human judgements in dialogue systems. 
Despite the initial insights provided, several limitations lie in these prior meta-evaluation studies: (1) most of them focused only on SRST: they 
have measured system performance in the context of question-and-answer dialogue systems where it is not possible to either provide multiple responses in a single turn, or converse throughout multiple rounds (turns).
(2) these metric meta-evaluation studies have only examined \emph{fidelity} (i.e.,~whether the metric agree with users' assessments or preferences) and for a limited set of metrics. 

\begin{figure}[t]
    \centering
    \includegraphics[width=10cm]{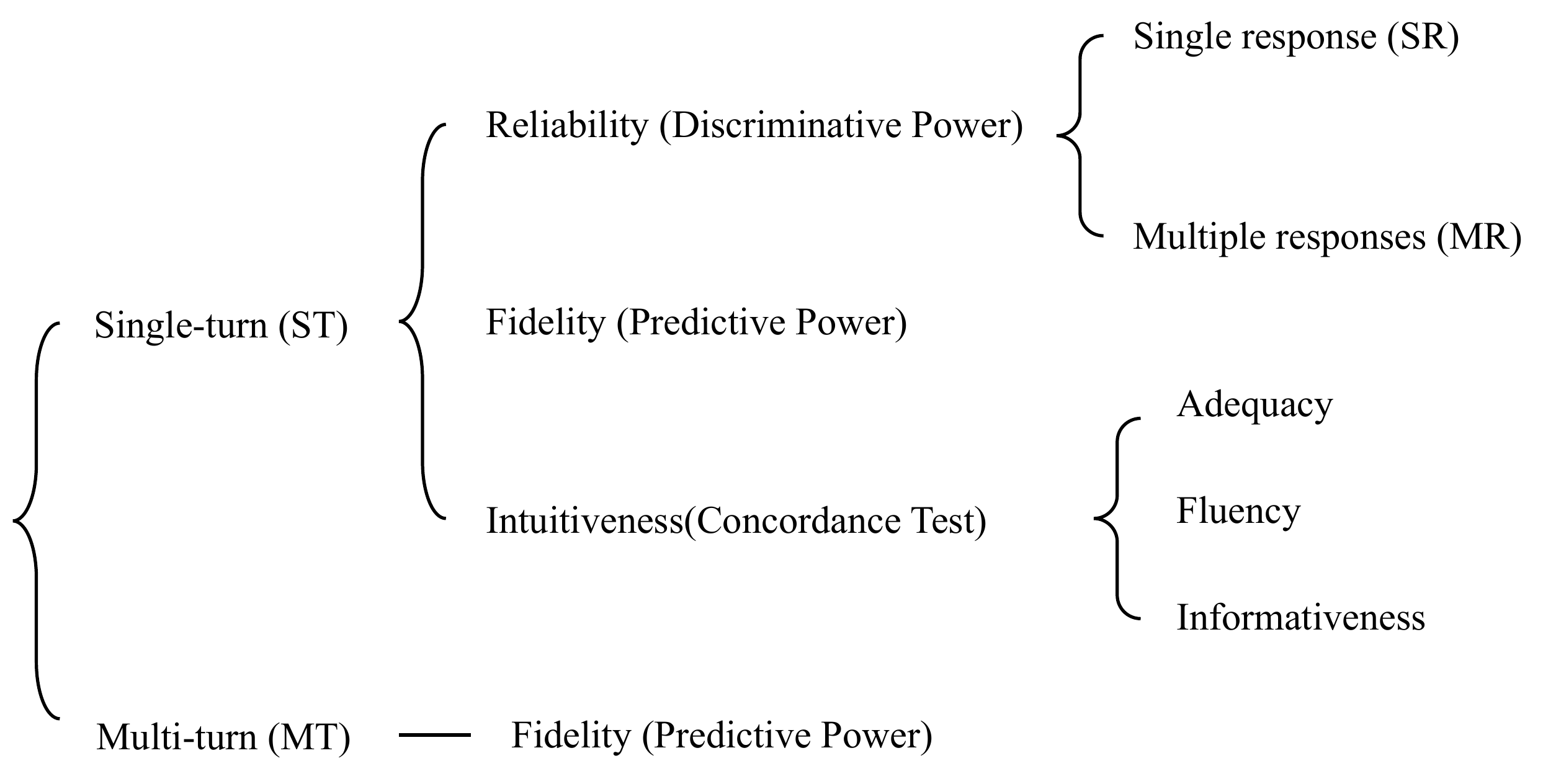}
    \caption{Overview of Our Meta-evaluation Framework.}
    \label{fig:intro_framework}
\end{figure}

To address the above limitations, we systematically 
investigate the performance of a set of representative existing metrics, including dialogue metrics and IR metrics, in both single-turn and multiple-turn conversational search environments. 
The comprehensive meta-evaluation framework we adopt in our paper is shown in Figure~\ref{fig:intro_framework}. 
In contrast to previous studies, our meta-evaluation work focuses on three different unique perspectives:\
\begin{itemize}
\item \emph{Reliability}: the ability of a metric to detect ``actual'' performance differences as opposed to those observed by chance. In this paper, reliability is measured using discriminative power \cite{sakai2006evaluating}. We use the randomised Tukey’s Honestly Significant Differences test \cite{carterette2012multiple} because, as shown by \citet{sakai2012evaluation}, this test is less likely to find significant differences that are not ``actual''.
\item \emph{Fidelity}: the ability of a metric to measure what it intends to measure and agree with ultimate user preferences. We adopt predictive power \cite{sanderson2010user} to capture the extent of a given evaluation metric's ability to predict a user's preference.
\item \emph{Intuitiveness}: the ability to capture any property deemed important in a metric. We quantitatively measure the preference agreements using the concordance test \cite{sakai2012evaluation} of a given conversational metric with a ``basic'' metric that captures specific key properties that a good conversational search system should satisfy. In our paper, we choose \emph{adequacy}, \emph{fluency} and \emph{informativeness}, three crucial indicators that influence user preference \cite{papineni2002bleu, liu2016not}.
\end{itemize}
All three of these perspectives are crucial to systematically understand a metric's ability to effectively evaluate search \cite{sakai2013metrics}. 
In addition, different from prior studies \cite{liu2016not, novikova2017we}, we move beyond SRST metrics: we perform meta-evaluation on (1)~MRST metrics that involve multiple responses in a single turn conversation, and (2)~multi-turn metrics adapted from existing search session-based measures \cite{liu2018towards, jarvelin2002cumulated, jiang2015understanding, jiang2016correlation}. Our meta-evaluation covers a wide and more comprehensive range of metrics in comparison to prior studies.
By conducting empirical experiments on two large test collections, we find that the performance of different metrics vary:
\begin{itemize}
    \item In terms of reliability, we find that METEOR \cite{banerjee2005meteor} and \llzy{BERTScore \cite{bert-score}} are able to perform robustly (i.e., having better discriminative power scores than most of other existing metrics) in both SRST and MRST scenarios. When considering multiple responses, RBP \cite{moffat2008rank} outperforms others in capturing the diminishing return of ranked responses.
    \item In terms of fidelity, similar to prior studies, we also find a weak correlation between existing metrics and human judgements (user preference). It is worth noting that embedding-based metrics are generally less predictive than word overlap-based metrics across different test collections. \llzy{Furthermore, we also observe the learning-based metric BERT-RUBER \cite{ghazarian2019better} can achieve good correlation with human preference in some specific collections, but its performance is not robust in different datasets. } \lllzy{This is because the training model part (i.e., the Unreferenced part) in BERT-RUBER is sensitive to the training collection, which further affects the whole performance of the metric.} 
    \item In terms of intuitiveness, we find that BLEU \cite{papineni2002bleu} and \llzy{METEOR \cite{banerjee2005meteor}} perform the best in capturing adequacy and informativeness whereas if fluency is the most important property to quantify, the \llzy{learning-based metric BERT-RUBER \cite{ghazarian2019better}} could be adopted \lllzy{because BERT-RUBER is more consistent with the automatic fluency gold standard (i.e., SLOR \cite{kann2018sentence})}. 
    \item When conducting multi-turn evaluation, those adapted search session-based metrics \lllzy{(e.g., sCG \cite{liu2018towards}, sDCG \cite{jarvelin2002cumulated})} are found to be moderately concordant with user satisfaction. \lllzy{We find that the Max strategy (i.e., to choose the highest relevance score in a session) can achieve the highest correlation with human satisfaction annotations. }
\end{itemize}
Overall, when considering all three perspectives, we find that METEOR \cite{banerjee2005meteor} is comparatively speaking the best metric.

\begin{figure}
    \centering
    \includegraphics[width=\textwidth]{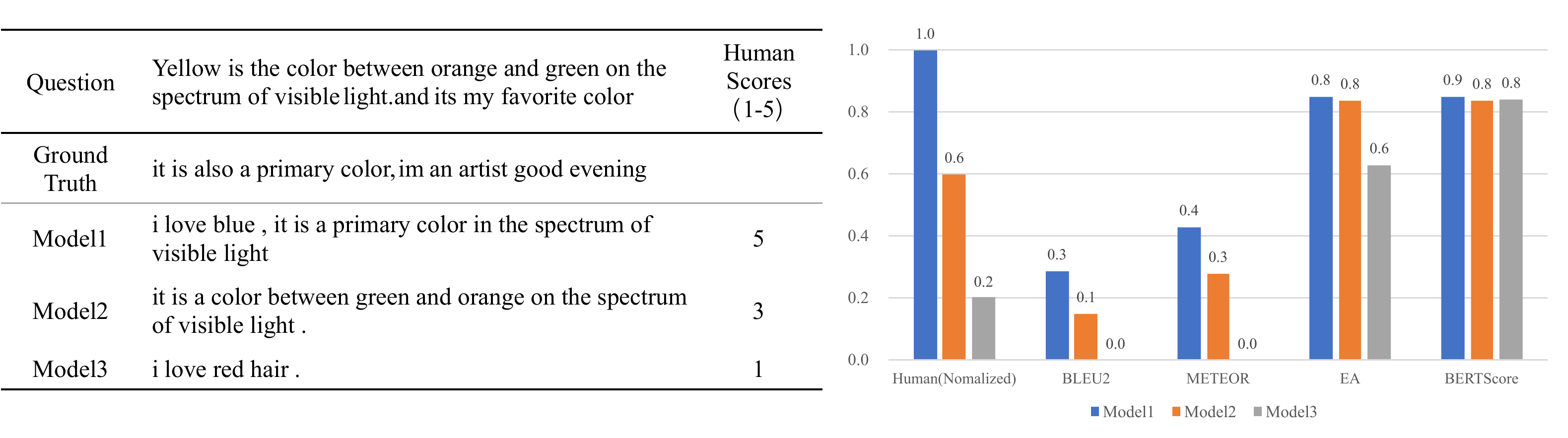}
    \caption{An example of different metric scores in conversational search evaluation. In human scores, `5' means the response is highly related to the question and `1' means the response is totally irrelevant to the question.}
    \label{fig:example_metrics}
\end{figure}

\lllzy{Figure ~\ref{fig:example_metrics} shows an example of different metrics for evaluating different models' responses. We can see that metrics may show different extents of correlation with human annotations. For example, BLEU2 and METEOR can correctly distinguish the difference between these three models' responses, and highly correlate with the variation of human annotations. Conversely, BERTScore values for these three responses are very similar and struggles to distinguish between them, partly because BERTScore considers more related words and misuses these contextual words (such as `green' in Model 2 and `red' in Model 3, which are close to the word `color') to estimate the similarity between responses and ground truth. This example shows that metrics may perform differently in the conversational search scenario and we consider it necessary to further investigate how well such metrics perform.} 





The main contributions of our work are three fold:
\begin{itemize}
    \item We present a systematical meta-evaluation of conversational search metrics. To our best knowledge, this work is the first to comprehensively study the reliability and intuitiveness of existing conversational search metrics.
    \item Moving beyond single-turn responses like a QA system, we meta-evaluate conversational search metrics for other presentation strategies with multiple responses within single-turn conversations.
    \item As a first step, we adapt search session based metrics for multi-turn conversational search evaluation. 
\end{itemize}



\S\ref{sec:related work} presents previous work, briefly introducing the methods of meta-evaluation and satisfaction evaluation. The main metrics and test collections used in this paper are described in \S\ref{sec:metric}. \S\ref{sec:metaevaluationsingleturn} focus on the meta-evaluation, including discriminative power \S\ref{sec:discriminativepower}, predictive power \S\ref{sec:predictivepower} and intuitiveness \S\ref{sec:intuitiveness}, in the single turn scenario and systematically introduce the methodology, experiment settings and experiment results. A brief summary of single turn meta-evaluation findings is presented at \S\ref{sec:summaryofsingleturn}. \S\ref{sec:multipleturn} shows the meta-evaluation in the multi-turn environment. A discussion and conclusion of this work are presented in \S\ref{sec:discussion} and \S\ref{sec:conclusion}.

%% file: content/2relatedwork.tex
\subsection{Evaluation of Conversational Search}


\lllzy{Evaluating conversational search systems is an open problem. Given that conversational searches usually start with an indefinite goal and lack an uniform structure like traditional search systems, it is difficult to find appropriate features to measure the quality. Depending on the specific characteristics of evaluation scenarios, two types of evaluation approaches are widely used in the conversation evaluation: offline methods and online methods. In this section, we discuss the general metrics which are applied in these two above methods, respectively. }

\lllzy{\textbf{Offline methods.} Offline methods focus on static collections where the candidate responses and ground-truth (ideal responses) in the conversation are already known. The basic idea of offline methods is to evaluate the appropriateness of responses by comparing the candidate system responses with the ideal ones, which are usually generated by humans. Here, appropriateness encapsulates many finer-grained concepts, such as coherence, relevance, and correctness \cite{deriu2020survey}. There are three main types of metrics in offline methods: word overlap-based metrics, word embedding-based metrics and learning-based metrics.}
\lllzy{
\begin{itemize}
    \item \emph{Word overlap-based metrics.} These metrics are originally proposed by the machine translation field. The basic idea of these metrics is to count the number of the overlap words between candidate responses and the ground truth. Due to their simple algorithms and easy deployment, they have become a popular choice of metrics for evaluating conversational systems. Many dialogue related competitions, such as DSTC-8 \cite{kim2019eighth}, choose this type of metrics as standard metrics. Popular metrics such as BLEU \cite{papineni2002bleu} and METEOR \cite{banerjee2005meteor} are widely used to evaluate the adequacy of a response. However, some previous studies \cite{liu2016not,novikova2017we} indicate that these overlap-based metrics weakly correlate with human judgements.
    \item \emph{Word embedding-based metrics.} Exact word matching in word overlap-based metrics has a major limitation: they can not consider the connection between words that are similar topically. Therefore, embedding-based metrics are proposed to address this issue. Popular metrics such as Greedy Matching \cite{rus2012comparison}, Vector Extrema\cite{forgues2014bootstrapping} and BERTScore \cite{bert-score} are also applied in dialogue evaluation \cite{lan2019talk, mehri2020usr}.
    \item \emph{Learning-based metrics.} These metrics are based on predictive models inspired by the PARADISE framework \cite{deriu2020survey}. The basic idea of these metrics is to use a training model to fit appropriateness ratings by human judges. The features adopted in training models can be the semantic features of ground truth or the questions (or queries). For example, ADEM \cite{lowe2017towards} is a evaluation method using a recurrent neural network to predict the adequacy scores of utterances. \citet{tao2018ruber} proposed a mixed evaluation method combining referenced and unreferenced metrics. In their unreferenced part, they train a recurrent neural network to predict the appropriateness of a response with respect to a question. Since the model training process is significantly influenced by the quality of the train set, the performance of these metrics may change considerably in different dateset. 
\end{itemize}
}

\lllzy{\textbf{Online methods.} To the contrary, online methods mainly concentrate on users' behavior and feedback when interacting with systems in real time. Different from offline methods, online evaluation usually does not have ground truth for each utterance due to the real time interaction. Therefore, user experience is the key concern in online evaluation work. Previous work has proposed several aspects to measure user experience, such as user engagements \cite{sano2016prediction} and satisfaction \cite{kiseleva2016predicting}. Especially, user satisfaction has become one important indicator that reflect the quality of dialogues. The concept of satisfaction was first proposed by \citet{su1992evaluation}. \citet{kelly2009methods} further give a detailed definition as ``satisfaction can be understood as the fulfillment of a specified
desire or goal''.  Many prior studies have presented methods for satisfaction prediction for intelligent assistants, such as \cite{kiseleva2016predicting, kiseleva2016understanding, hashemi2018measuring}. The basic idea of these methods is to construct a predictive model based on user interaction behavior signals or semantic features, and estimate a score of utterance which is close to human real judges. Due to the lack of user interaction in our collections, we do not meta-evaluate the satisfaction methods in this paper. However, since satisfaction is a key concern in conversation evaluation, we also consider the correlation between the selected metrics and user annotated satisfaction scores. }

\subsection{Meta-evaluation Methods}


\lllzy{A metric is often designed for measuring the performance of search systems from one or several specific perspectives. Therefore, different metrics concentrate on various aspects and have their own strengths and weaknesses. Meta-evaluation is an available evaluation tool that can compare the performance of metrics from different aspects, which include reliability \cite{sakai2007reliability}, sensitivity \cite{radlinski2010comparing, sakai2005effect}, fidelity \cite{sanderson2010user, liu2016not}, agreement \cite{turpin2006user}, intuitiveness \cite{sakai2012evaluation, zhou2013reliability}, unanimity \cite{sakai2019diversity} and so on. In this paper, we focus on reliability, fidelity, and intuitiveness because these three aspects are often the key concerns in metric design \cite{liu2016not, sakai2012evaluation, zhou2013reliability}. As sensitivity and agreement are also important indicators for meta-evaluation, these should be further considered in future work when suitable approaches have been identified.}

\lllzy{For reliability, there are many different ways to evaluate the metrics, such as those proposed by  \citet{sakai2007reliability} and \citet{radlinski2010comparing}. It is worth noting that discriminative power only `compares the statistical stability of measures by means of obtaining the p-value for every system pair', while `Unanimity quantifies how a measure agrees with the SERP preferences according to all other measures' \cite{sakai2019diversity}. This might be a limitation of discriminative power. In this paper, we adopt \emph{discriminative power} to evaluate the reliability aspect of metrics because discriminative power does not need additional human annotations and has been demonstrated by many previous works \cite{sakai2006evaluating, sakai2012evaluation, zhou2012evaluating, sakai2019diversity}. }

\lllzy{For fidelity, one popular way is to compare the correlation rate between human annotation scores and the metric scores, such as those proposed by \citet{liu2016not} and \citet{novikova2017we}. In this study, we choose to use \emph{predictive power}, which measures the agreement between user preferences and metrics when presenting a pair of different responses, to evaluate metrics in terms of fidelity. Many prior studies have proved its 
suitability for measuring the fidelity of metrics in meta-evaluation, such as \cite{sanderson2010user, zhou2012evaluating, sakai2012evaluation, chen2017meta}. Since it is relatively easy to extract response pairs and acquire users' preference of these pairs from the original collections (the details are shown in \S\ref{sec:collection}), predictive power is a good choice to measure the fidelity of metrics in our experiment.}

\lllzy{In terms of intuitiveness, one popular method is to use concordance test to capture the important property with gold standard metric, as shown by \citet{sakai2012evaluation}, \citet{zhou2013reliability} and \citet{chen2017meta}. Following these previous works, we also adopt similar settings to systematically investigate metrics in terms of \emph{adequacy}, \emph{fluency} and \emph{informativeness}.}

\lllzy{The ultimate goal of our work is to compare the performance of existing metrics from different perspectives. Few studies have thoroughly investigated the performance of metrics in the conversational search scenario. To our knowledge, our work is the first to systematically meta-evaluate both IR metrics and dialogue metrics for conversational search. Different from previous studies that only use a single aspect of meta-evaluation method, multiple different meta-evaluation methods are adopted in this work. We use discriminative power and concordance test to test the reliability and intuitiveness of metrics, and measure the fidelity of metrics based on predictive power.}

%% file: content/3metrics.tex

\lllzy{In this section, we introduce the key elements of our meta-evaluation, including the metrics (\S\ref{sec:overviewmetric}-\S\ref{sec:multipleturnmetric}), the datasets (\S\ref{sec:collection}) and the systems (\S\ref{sec:systemruns}).}
%
Our meta-evaluation (\S\ref{sec:metaevaluationsingleturn} and \S\ref{sec:multipleturn}) follows the following process using those key elements:
(1) We first collected two aforementioned human-to-human conversation datasets (\S\ref{sec:collection}), including utterances and human annotations. (2) Then we adopt rule-based methods to select reliable responses and regard them as the reference (ground-truth) responses. (3) After that, different systems (\S\ref{sec:systemruns}) are used to generate a response for each question. The quality of each response will be measured by the different evaluation metrics (\S\ref{sec:overviewmetric}-\S\ref{sec:multipleturnmetric}). (4) Finally, the performance of the metrics are meta-evaluated (\S\ref{sec:metaevaluationsingleturn}) from various aspects (i.e., reliability, fidelity and intuitiveness) by comparing the overall performance of different systems which are measured by these metrics.

\subsection{Overview of Selected Metrics} \label{sec:overviewmetric}

A wide range of metrics have been proposed to evaluate search and dialogue systems. 
Table ~\ref{tab:metric} 
provides an overview of the metric types and all the metrics we select in our meta-evaluation work. 
Since it is impractical to test all existing dialogue and IR metrics, we select these metrics according to two criteria: (1) \emph{popularity}: the metric 
has been widely used in academic research and in industry; (2) \emph{interpretability}: the metric can be directly interpreted and have simple structures, such that they can be expanded easily. We also make sure that all the metrics that were meta-evaluated in prior studies \cite{zhou2013reliability, liu2016not} are included in our work for comparison purposes. 
%
%
%
%
%
\begin{table*}[t] \footnotesize
\caption{Metrics meta-evaluated in this paper}
\begin{tabular}{|c|ccc|cl|}
\hline
 & \multicolumn{3}{c|}{Single turn (ST)} & \multicolumn{2}{c|}{\multirow{2}{*}{Multiple turn (MT)}} \\ \cline{2-4}
 & \multicolumn{1}{c|}{Single response (SR)} & \multicolumn{2}{c|}{Multiple responses (MR)} & \multicolumn{2}{c|}{} \\ \hline
\multirow{3}{*}{Word overlap-based} & \multicolumn{1}{c|}{BLEU \cite{papineni2002bleu}} & \multicolumn{1}{c|}{BLEU \cite{papineni2002bleu}} & \multirow{7}{*}{\begin{tabular}[c]{@{}c@{}}nDCG \cite{jarvelin2002cumulated} \\ ERR \cite{chapelle2009expected} \\ RBP \cite{moffat2008rank} \end{tabular}} &  &  \\
 & \multicolumn{1}{c|}{METEOR \cite{banerjee2005meteor}} & \multicolumn{1}{c|}{METEOR \cite{banerjee2005meteor}} &  &  &  \\
 & \multicolumn{1}{c|}{ROUGE \cite{lin2004rouge}} & \multicolumn{1}{c|}{ROUGE \cite{lin2004rouge}} &  &  &  \\ \cline{1-3}
\multirow{3}{*}{Embedding-based} & \multicolumn{1}{c|}{EA \cite{serban2017hierarchical}} & \multicolumn{1}{c|}{EA \cite{serban2017hierarchical}} &  &  &  \\
 & \multicolumn{1}{c|}{SCS \cite{sidorov2014soft}} & \multicolumn{1}{c|}{SCS \cite{sidorov2014soft}} &  &  &  \\
 & \multicolumn{1}{c|}{BERTScore \cite{bert-score}} & \multicolumn{1}{c|}{BERTScore \cite{bert-score}} &  &  &  \\ \cline{1-3}
\multirow{1}{*}{Learning-based} & \multicolumn{1}{c|}{BERT-RUBER \cite{ghazarian2019better}} & \multicolumn{1}{c|}{BERT-RUBER \cite{ghazarian2019better}} &  &  &  \\ \hline
\multirow{3}{*}{Session-based} &  &  & \multicolumn{1}{l|}{} & sCG \cite{liu2018towards} & \multicolumn{1}{c|}{SWF \cite{liu2018towards}} \\
 &  &  & \multicolumn{1}{l|}{} & sDCG \cite{jarvelin2002cumulated} & \multicolumn{1}{c|}{Max \cite{liu2018towards}} \\
 &  &  & \multicolumn{1}{l|}{} & sDCG/q \cite{jiang2016correlation, jiang2015understanding} & \multicolumn{1}{c|}{Min \cite{liu2018towards}} \\ \hline
\end{tabular}
\label{tab:metric}
\end{table*}
\lllzy{As shown in Table ~\ref{tab:metric}, four types of metrics are considered in this paper: word overlap-based, embedding-based, learning-based, and session-based metrics}.
\lllzy{
\begin{itemize}
    \item For STSR, we consider word overlap-based, embedding-based and learning-based metrics. It is worth noting that all these metrics are reference-based metrics. They are calculated by comparing candidate responses (generated by systems) and ground truth (reference response formulated by human responder).
    \item For STMR, except for the three aforementioned metrics, we also consider ranking-based metrics (i.e., nDCG \cite{jarvelin2002cumulated}, RBP \cite{moffat2008rank} and ERR \cite{chapelle2009expected}). These metrics use different 
    strategies to calculate the relevance of the entire ranked list of responses. Note that the relevance of each response in the result lists is calculated by STSR metrics based on the question and the response. 
    \item For MT scenarios, we focus on the performance of session-based metrics. Note that the relevance between a question and a response in each turn is also based on STSR metric scores. After that, we further combine these single-turn scores to calculate the final scores of the entire multi-turn dialogues based on different existing session-based IR metrics. 
\end{itemize}}


We next describe all our selected metrics in details, including single turn (\S ~\ref{sec:singleturnmetric}) and multi-turn metrics (\S ~\ref{sec:multipleturnmetric}).

\subsection{Single-turn Metrics} \label{sec:singleturnmetric}
\subsubsection{Word overlap-based metrics} \label{sec:wordoverlapmetric}
\

\noindent \textbf{BLEU} \cite{papineni2002bleu} is one of the most popular automatic metrics adopted for evaluation of dialogue systems. 
\lzy{Many previous studies use BLEU as a gold standard metric to evaluate the performance of dialogue models, especially generative models, such as those by \cite{ritter2011data, sordoni2015neural, tian2017make, song2016two}. Some studies \cite{galley2015deltableu, liu2016not}, however, show a weak correlation between BLEU scores and human annotations (e.g., satisfaction) and further suggest that BLEU might be inadequate for reflecting the real quality of dialogues. Nevertheless, BLEU is widely adopted due to its simplicity.} 
BLEU is a precision-oriented metric and focuses on the matches of the word sequences, or n-grams in both the proposed responses and ground truth. We denote the n-gram precision score for the whole test collection as $P_n$. Equation ~\ref{equ:bleu} shows the calculation of $P_n$, where $Cnt(n\mhyphen gram)$ is the number of n-gram occurrences in a candidate response $r$, and $Cnt_{clip}(n\mhyphen gram)$ is \lzy{the maximum number of n-grams co-occurring in a candidate response $r$ and a reference (ground-truth $g$)}. Considering a BLEU score may fail to evaluate any short responses that only contain the highly confident n-grams, BLEU-N provides a way to address this weakness. 

\begin{equation}
    Prec_{n} = \frac{\sum_{r\in{candidates}}\sum_{n\mhyphen gram\in{r}} Cnt_{clip}(n\mhyphen gram)}{\sum_{r\in{candidates}}\sum_{n\mhyphen gram\in{r}} Cnt(n\mhyphen gram)}
    \label{equ:bleu}
\end{equation}
Equation ~\ref{equ:bleun} shows the computation of BLEU with a brevity penalty, where N is the the maximum length of n-grams, and BP is the brevity penalty (shown in Equation ~\ref{equ:bleu_bp}) based on the length of the candidate $r$ and the ground truth $g$. In practice, BLEU metric commonly uses N = 4 and $W_n$ is typically set to 1/N. 

\begin{equation}
    BP = \begin{cases}1 & if|r|>|g|\\e^{1-|g|/|r|} & if|r| \leq |g|\end{cases}
    \label{equ:bleu_bp}
\end{equation}

\begin{equation}
    BLEU\mhyphen N = BP \cdot \exp{\sum_{n=1}^N W_n\log{Prec_n}}
    \label{equ:bleun}
\end{equation}
Note that BLEU is designed for large-scale test corpora and it is not considered reliable when used on individual sentences \cite{liu2016not}. In addition, BLEU also lacks synonym matching and is unable to detect the orders of multiple words. Despite these shortcomings, BLEU is still one of the most widely used automatic metrics in conversational searches.  

\noindent \textbf{METEOR.} The primary goal of METEOR \cite{banerjee2005meteor} is to address several observed shortcomings in BLEU. Unlike BLEU, which only calculates n-gram precision, METEOR take both precision and recall into account, by computing a parameterized harmonic mean. Equation ~\ref{equ:meteor} shows the computation of $METEOR$, where precision $Prec$ is a number of matches divided by the unigram number in the proposed responses, and recall $Rec$ is the number of matches divided by the ground truth. Note the matches of METEOR includes exact matching, stem matching, and WordNet synonyms matching. That means METEOR can improve the ability to handle the variability of responses due to the incorporation of synonym and stem matching.

\begin{equation}
    METEOR = (1-Pen)\cdot \frac{Prec\cdot Rec}{\alpha Prec + (1-\alpha )Rec}
    \label{equ:meteor}
\end{equation}

\begin{equation}
    Pen = 0.5 * (\frac{|n_{\lzyfinal{chunks}}|}{|n_{unigram}|})^{3}
    \label{equ:meteor_pen}
\end{equation}
METEOR also provide a penalty $Pen$ to reduce the effect of short sequences of consecutive matches, namely, \lzyfinal{chunks}. With the alignment between the candidate and ground truth, $Pen$ is calculated as Equation ~\ref{equ:meteor_pen}, where $n_{\lzyfinal{chunks}} $ is the fewest possible number of \lzyfinal{chunks} and $n_{unigram}$ is the number of unigrams matched between candidate responses and ground truth.
Unlike BLEU, METEOR does not penalize long responses and is able to identify equivalences between candidate responses and ground truth \cite{olive2011handbook}.

\noindent \textbf{ROUGE} \cite{lin2004rouge} is a set of recall-oriented metrics used for automatic evaluation of summaries.
ROUGE-L is widely applied to the evaluation of dialogue systems \cite{zhang2018generating}, which is a Long Common Subsequence (LCS) based F-measure. Its uses in-sequence matches to indicate the word order, and does not require consecutive co-occurrences between the candidate $r$ and the ground-truth $g$.
Equation ~\ref{equ:rouge} shows the calculation process of recall (Rec), precision (Prec) and F-score (F):
\begin{equation}
    Rec = \frac{LCS(X, Y)}{Length_X};
    Prec = \frac{LCS(X, Y)}{Length_Y};
    F_{lcs} = \frac{(1 + \beta^2)Rec \cdot Prec}{Rec + \beta^2Prec},
    \label{equ:rouge}
\end{equation}
where X denotes a reference sentence and Y denotes a candidate sentence.
LCS(X,Y) is the length of a longest common subsequence of X and Y, and $F_{lcs}$ is the ROUGE-L score. Generally, $\beta$ is set to a very big number, so ROUGE-L metric only considers recall (Rec). In other words, ROUGE-L is still a recall-orient evaluation metric.

\subsubsection{Word Embedding-based metrics} \label{sec:wordembeddingmetrics}
\ 



\noindent \textbf{Embedding Average (EA)} has been widely used in dialogue system evaluation \cite{serban2017hierarchical, jin2018explicit}. Previous studies provide many methods to evaluate sentence-level similarity based word embedding, such as greedy matching \cite{rus2012comparison}, embedding average \cite{foltz1998measurement, landauer1997solution, mitchell2008vector}, and vector extrema \cite{forgues2014bootstrapping}. In this paper, we only select the embedding average metric \lzy{because the differences among these three metrics' performance are small \cite{liu2016not}.} 
The EA metric calculates sentence-level embedding by averaging the vector of the individual word in a sentence: 
\begin{equation}
    \bar{e_s} = \frac{\sum_{w\in s} e_w}{N_s}
    \label{equ:ea}
\end{equation}
where $e_w$ is the word vector of each token in a sentence $s$, and $N_s$ is the number of the tokens in this sentence. Following previous work \cite{liu2016not}, we adopt the cosine similarity to compare the proposed response $r$ with the ground truth $\hat{g}$, as shown in Equation ~\ref{equ:eacos}.
\begin{equation}
    EA = \cos(\bar{e_r}, \bar{e_g})
    \label{equ:eacos}
\end{equation}

\noindent \textbf{Soft Cosine Similarity (SCS)}  \cite{sidorov2014soft} is a sentence-level metric using word embedding, which is also used in the dialogue community \cite{charlet2017simbow}. Unlike standard cosine similarity, soft cosine similarity further considers word-level relations by creating a word relation matrix based on similarity measures (e.g., synonymy). To compare a candidate response with the ground truth, we first compute the cosine similarity of all word pairs using the embedding of each word, and construct a similarity matrix for the whole corpus. Note that we only use FastText \cite{joulin2016fasttext} to learn text representations and convert the word to the vectors. We leave the appropriate tuning of embedding methods for future work.
Equation ~\ref{equ:softcosine} shows how to calculate the SCS metric:
\begin{equation}
    soft\mhyphen cos(S_g, S_r) = \frac{\sum\sum_{i,j}^N m_{i,j}w_{g,i}w_{r,j}}{\sqrt{\sum\sum_{i,j}^N  m_{i,j}w_{g,i}w_{g,j}}\sqrt{\sum\sum_{i,j}^N  m_{i,j}w_{r,i}w_{r,j}}}
    \label{equ:softcosine}
\end{equation}
where $m_{i,j}$ is the similarity score between word $i$ and word $j$, \llzy{which in this paper is calculated by using cosine similarity}; $w_{g,i}$ means \llzy{the embedding vector of} word $i$ in the ground truth; and $w_{r,j}$ is \llzy{the embedding vector of} word $j$ in the proposed response. With this metric, the similarity between two sentences is not null if these two sentences share related words, even if they have no common words. 

\noindent \textbf{BERTScore} \cite{bert-score} is one of the latest proposed \llzy{embedding-based evaluation metrics for text generation, which is based on BERT contextual embeddings}. We choose it as a candidate metric because: (1) BERTScore can achieve a good correlation with human judgements in both machine translation and image captioning tasks \cite{bert-score}; and (2) it has been widely used in open dialogue system evaluation \cite{lan2019talk, mehri2020usr}. 
%
BERTScore consists of four main parts: token representation, similarity measure, score calculation and importance weighting. Firstly, contextual embeddings (i.e., BERT \cite{devlin2018bert}) are used to represent the tokens in the input sentences. Then these pre-normalized token vectors allow for a soft measure of similarity. After that, the BERTScore can be calculated by the token-level matching, including recall, precision and F1 measure. They use greedy matching to maximize the matching similarity score. 
According to the results of Zhang et al. \cite{bert-score}, we regard the F1 measure, which generally achieves a better performance than precision and recall, as the standard score of BERTScore. Equation ~\ref{equ:bertscore} shows the calculation of recall ($Rec$), precision ($Prec$), and F1 score ($F$):
\begin{equation}
    Rec=\frac{1}{|e_g|}\sum_{e_i \in e_g}{\max_{e_j \in e_r}{e_i^Te_j}};
    Prec=\frac{1}{|e_r|}\sum_{e_j \in e_r}{\max_{e_i \in e_g}{e_i^Te_j}};
    F = 2\frac{Prec \cdot Rec}{Prec+Rec}
    \label{equ:bertscore}
\end{equation}
where $e_g$ denotes the contextual embeddings of the ground truth and $e_r$ is the embeddings of output responses. 

\subsubsection{Learning-based metrics} 
\ 


\noindent \lllzy{\textbf{BERT-RUBER} \cite{ghazarian2019better} is one of the state-of-the-art learning-based evaluation metrics for open-domain dialogue systems. Like other typical learning-based metrics (e.g., ADEM \cite{lowe2017towards}, RUBER \cite{tao2018ruber} and PONE \cite{lan2020pone}), the basic idea of BERT-RUBER is to train a score model to evaluate the response based on its context. Since BERT-RUBER outperforms RUBER and ADEM \cite{ghazarian2019better}, and it is difficult to interpret complex learning-based metrics like PONE \cite{lan2020pone}, we choose BERT-RUBER as the sole representative learning-based metric in this paper.} 
\llzy{BERT-RUBER consists of an unreferenced metric and a referenced metric. The unreferenced metric is a training model, which aims to predict the relevance between responses and the given queries (or questions). Note that when evaluating a response, the unreferenced metric part is calculated without any references or ground truth for the given question. It is a predictive model whose input are candidate response and the questions, which is trained based on a collection of pairs of question and reference response.} \llzy{Meanwhile, the referenced metric computes the cosine similarity between responses and references (ground-truth) base on BERT embeddings. Then, the final BERT-RUBER score combine those two metrics above by heuristics methods(e.g., min, max).}
\llzy{Since the unreferenced metric part is a learning-based model, the performance of BERT-RUBER can be influenced by the training dataset. In order to achieve good performance of the metrics, we train and tune the unreferenced model base on the specific dataset which we use.
Following previous work \cite{ghazarian2019better}, we also use 2 layers of bidirectional gated recurrent unit with 128-dimensional hidden unit and apply three layers for MLP (Multilayer Perceptron Network) with 256, 512 and 128-dimensional hidden units. Learning rate decay is applied when no improvement was observed on validation data for five consecutive epochs. }

\subsubsection{Ranking-based metrics} \label{sec:rankingbasedmetric}
\ 

In addition to providing a single response, \citet{boussaha2019deep} and \citet{song2018ensemble}
use a list-like response as an alternative way for conversational search systems to meet users' information needs. In this paper, we also take ranking-based metrics into account. We consider 
evaluating the quality of the answer list as a whole. Although recall is a popular metric, which has been used to evaluate dialogue models \cite{lowe2016evaluation, lowe2015ubuntu, zhou2016multi, wu2016sequential, yang2018response, dinan2018wizard}, it is still inadequate to evaluate the quality of this kind of response because it assumes relevance is a binary judgement, and does not consider the ranking position. Therefore, in order to incorporate graded relevance, we choose a small number of popular ranking-based metrics, such as nDCG@k \cite{jarvelin2002cumulated}, as the basic metric for MR scenario.
%
\lzy{It is worth noting that these ranking-based metrics cannot be adopted directly in the MR scenario because of the lack of 
corresponding relevance labels. In MR scenarios, we assume the search system returns the top 5 candidate responses as a whole list to answer the user's question in each turn. In order to align the experimental settings, here we treat each response in the list as a document.}
\lzy{In terms of relevance judgements, since single-response `metrics' (e.g., BLEU, METEOR) can be often regarded as crude indicators for users' preferences \cite{papineni2002bleu, banerjee2005meteor}, we replace the relevance judgements in the ranking-based metrics by calculating the single-response `metric' score. Although these metric scores are not strictly equal to the relevance judgments, these scores could reveal the similar trend of user preference to some extent. Thus, we can use existing ranking based IR metrics as follows: `response' is treated `document', whereas `SR metric score' is used as `relevance judgements'.}


\noindent \textbf{nDCG.} nDCG@k \cite{jarvelin2002cumulated}, or Discounted Cumulative Gain, is a ranking-based measurement that evaluates the performance of top $k$ by cumulating the gain for each position $i$. The equation ~\ref{equ:ndcg} shows the definition of this metric.

\begin{equation}
    nDCG@k = \frac{\sum_{i=1}^k{(2^{R_i} -1) } / \log{(i+1)} }{\sum_{i=1}^{k'}{(2^{R'_i} - 1)} / \log{(i+1)}}
    \label{equ:ndcg}
\end{equation}
where $R_i$ is the graded relevance of the document at position $i$, and the entire metric score is normalized by an ideal ranked list k'. The basic assumption of this simple metric is that systems that put highly relevant documents at the top of the result list are better than systems that put highly relevant documents further down the ranking. 

\noindent \textbf{RBP.} Rank-biased precision (RBP) is also one rank-based metric, which `measures the rate at which utility is gained by a user working at a given degree of persistence' \cite{moffat2008rank}. Different from the basic assumption of DCG, RBP further considers the persistence parameter $p$ to describe the user browsing behavior. The RBP metric is computed as below:
\begin{equation}
    RBP=(1-p)\cdot\sum_{i=1}^d{R_i \cdot p^{i-1}}
    \label{equ:rbp}
\end{equation}
where $R_i$ is the degree of \lzy{relevance of the document at rank} $i$ and $p$ is the persistence or probability when users examine from one document in a ranked list to the next.
\llzy{It is worth noting that the RBP metric value is influenced by $p$ \cite{moffat2008rank, sakai2008information, sakai2019diversity}. Considering the ranking list in the multiple response scenario is similar to traditional search, we use both p=0.5 and p=0.7 in the experiment. We choose p=0.5 given that it does not over-reward top responses in the ranking \cite{moffat2008rank}. We would also like to evaluate whether the same conclusions as in \citet{sakai2019diversity} holds in the context of conversational search. We choose p=0.7 given it has been shown to closely approximate the estimated probability of examination from prior work\cite{chapelle2009expected}. We leave the tuning of the optimal $p$ of RBP for conversational search as future work.}


\noindent \textbf{ERR.}  Expected Reciprocal Rank (ERR) is another popular ranking-based retrieval metric, which allows for the expected reciprocal length of time to find a relevant document\cite{chapelle2009expected}. Similar to RBP, ERR also takes user browsing behavior into account but focuses on cascade behavior. The definition of ERR metric is shown as equation ~\ref{equ:err}:

\begin{equation}
    ERR = \sum_{r=1}^{n}{\frac{1}{r}\prod_{i=1}^{r-1}{(1-R_i)R_r}}
    \label{equ:err}
\end{equation}
where $R_i$ is a mapping from relevance grade of $i^{th}$ document to probability of relevance, and $R_r$ denotes the probability of relevance when the user stops at position $r$ based on the cascade model.

All the ranking-based metrics above should have relevance annotations to calculate the scores. 
As aforementioned, we resort to using single-response metric to define the relevance gain. The relevance score $R_i$ is calculated by comparing ground-truth answer for all ``utterances''. Here we present the main steps in the calculation of the relevance score.

\begin{itemize}
    \item For nDCG,  $R_i = M(r, g)$;
    \item For RBP, $R_i = M(r, g)$;
    \item For ERR, $\displaystyle R_i = \frac{2^{M(r, g)}-1}{2^{M_{\max}{(r, g)}}}$
\end{itemize}
where $M(r, g)$ denotes a specific metric score (e.g., BLEU, METEOR) which is calculated on the response $r$ and the ground truth $g$. 
We minimize the modification of these metrics' original construction, adopt the most popular version, and only replace relevance labels with the single turn metric scores directly.

\subsection{Multiple-turn Metrics} \label{sec:multipleturnmetric}

\lzy{
As an interactive retrieval system, conversational search also shares some common features with `classical' search systems. 
Both conversational search and search sessions allow users to interact with systems in multiple rounds. }
In this section, we select session-based metrics to evaluate multi-turn conversational search.
We want to note that there are some important differences between session search and conversational search, which may limit the application of the session-based metrics in conversational search environments.
\lzy{First, there are no fixed set of documents for judging in conversational search. The responses (or retrieval documents) may change dynamically while users shift utterances during the dialogue. From this perspective, it is difficult to exploit judgments for the evaluation of conversational search.}
\lzy{Second, }the forms of search result are more flexible in conversational search. Conversational search systems can provide either single text response or multi-rankings during different rounds, while the session-level search often present search result list. 
%
%
\lzy{Considering these important differences, we assume that we have already known the sequence of dialogues with ground-truth of each round, and the set of responses in each dialogue is fixed. Moreover, we further assume users prefer to firstly trigger the dialogues and systems only provide responses according to the previous utterance. Following these two assumption, we define one conversational session as $
\left\{ q_1, r_1, q_2, r_2, q_3, r_3,..., q_i, r_i \right\}$. Let $q_i$ denote the utterance or questions in the $i^{th}$ turn in a session, and $rel(q_i, r_i)$ denote the relevance gain for the candidate response $r_i$ of the utterance $q_i$, which is calculated on single-turn metrics with ground truths. Then, we calculate the relevance gain of each turn separately and equation ~\ref{equ:gain} shows the calculation method of the gain of each turn. }

\begin{equation}
    gain(q_i,r_i) = 2^{rel(q_i, r_i)} - 1
    \label{equ:gain}
\end{equation}
%
In this paper, we choose sCG\cite{liu2018towards}, sDCG \cite{jarvelin2002cumulated}, sDCG/q\cite{jiang2016correlation, jiang2015understanding}, and session-level weighting functions \cite{liu2018towards} to perform the meta-evaluation in the multiple-turn scenario.\footnote{We note that session rank bias precision (sRBP) \cite{lipani2019user},
one of state-of-the-art session-based metrics, is not suitable for our experiment settings, where each `session' (i.e., each question-answer round) only contains one response. Since users only examine the first document but perform multiple rounds with the agent, all the users may be treated as patient users in this metric, and the \emph{balance} parameter could be 0 (sRBP score only the first document of every reformulation) and the \emph{persistence} parameter, which defines the persistence of users in continuing search, could become 1. Thus, all the sRPB scores of the sessions could be 0. Therefore, we do not discuss sRBP in our meta-evaluation experiment.}
\lzy{To fit the conversational search settings, we replace the original gain functions of each turn with our defined relevance gain function (Equation ~\ref{equ:gain}) in these session-based metrics.} 



\noindent \textbf{sCG.} This is a basic metric that is proposed in \cite{liu2018towards}. It sums up cumulative gain without query discount. 

\begin{equation}
    sCG = \sum_i^{|S|}gain(q_i,r)=\sum_i^{|S|}(2^{rel(q_i, r)}-1)
\end{equation}
where $|S|$ is the number of queries in the whole session.

\noindent \textbf{sDCG.} This metric is proposed in \cite{jarvelin2002cumulated}. This is a session-based discounted cumulative gain (sDCG) metric. The basic assumption of this metric is that results retrieved by later queries in a session provide less information gain because user have to take more effort when they reformulate queries. The main step of this metric is calculated as equation ~\ref{equ:sdcg} :

\begin{equation}
DCG = \sum_{i=1}^{|S|}\frac{gain(q_i,r_i)}{\log_2{(i+1)}} = \sum_{i=1}^{|S|}\frac{(2^{rel(q_i, r_i)} - 1)}{\log_2{(i+1)}}
\label{equ:dcg}
\end{equation}

\begin{equation}
sDCG = \sum_{i=1}^{|S|}\frac{DCG(q_i)}{\log_{bq}{(i+bq-1)}}
\label{equ:sdcg}
\end{equation}
where $DCG(q_i)$ is the discount cumulative gain for a single query's performance, and $bq$ denotes a parameter representing the extent to which the modeled user reformulates queries. Following the previous work \cite{jarvelin2002cumulated, kanoulas2011evaluating}, we use the recommend settings $bq = 4$ in our study.

\noindent \textbf{sDCG/q.} This is a normalized version of sDCG metric, which is proposed in \cite{jiang2016correlation, jiang2015understanding}. This metric simply use the number of queries to normalize the sDCG scores. Equation ~\ref{equ:sdcgq} shows the calculation of sDCG/q.

\begin{equation}
    sDCG/q = \frac{sDCG(S)}{n}
    \label{equ:sdcgq}
\end{equation}
where S represents the all queries in a session. Note that Kanoulas et al. \cite{kanoulas2011evaluating} propose another version of normalized sDCG metric, i.e., nsDCG. It assumes all the queries have an ideal ranked list and the session can achieve the ideal performance if each query have ideal ranked list. However, in our study, each question only has one response, which means we are not able to calculate the ideal DCG for each query. 

\noindent \textbf{Session-level Weighting Functions (SWF).} The basic idea 
\cite{liu2018towards} is to adopt different weight to each query position \lzy{(the rank of the query sequence in a session)} 
and sum up the satisfaction score of each query in a session:

\begin{equation}
    M=\sum_{i=1}^N w_i^* * s_i
    \label{equ:weightfun1}
\end{equation}
where $s_i$ is the user's satisfaction on query $i$. In our study, each question-answer (QA) pair does not have annotated satisfaction scores. Assuming the satisfaction correlates with the relevance of the responses, we use the gain score of each query instead of the satisfaction scores. $w_i^*$ can be calculated with Equation ~\ref{equ:weightfun2}.

\begin{equation}
    w_i^* = \frac{w_r}{\sum_{r=1}^{|S|}w_r}
\label{equ:weightfun2}
\end{equation}
where N denote the query number in a session, $w_r$ is the $r^{th}$ query's original weight which has been shown in Table ~\ref{tab:weightfun}.

\begin{table}[t] 

\caption{The r-th query's weight of session weighting functions \cite{liu2018towards}}
\begin{tabular}{ccc}
\toprule
Metric & $w_r(0 < r \leq N/2)$ & $w_r(N/2 < r \leq N)$ \\ \midrule
Decrease\_weight & 1/r & 1/r \\
Increase\_weight & r & r \\
Equal\_weight & 1 & 1 \\
Middle\_high & r & N+1-r \\
Middle\_low & 1/r & 1/(N+1-r) \\ \bottomrule
\end{tabular}
\label{tab:weightfun}
\end{table}

\subsection{Conversational Search Collections} \label{sec:collection}
Our study aims to meta-evaluate different metrics based on existing public datasets. Typically, existing datasets can be roughly grouped into three categories: machine-to-machine, human-to-machine, and human-to-human \cite{budzianowski2018multiwoz}. In this paper, we focus only on human-to-human corpora because human-to-human collections are more reliable and similar to the natural conversation \cite{budzianowski2018multiwoz}.
There are two common ways to collect this type of conversation data. One strategy is to build a dialogue system that can directly mimic the interaction process between users and agents, such as \cite{trippas2018informing}. \lzy{In the data collection process, one person plays an agent and the other person plays a user. The user role should ask questions to the agent role and finish the specified task.} The collections from this strategy usually have high quality due to the controllable experimental setup. 
The other way is to crawl dialogue data from the open web source, such as Twitter \cite{ritter2010unsupervised} and the forum of Ubuntu technical support \cite{lowe2015ubuntu}. These types of collection are typically noisy and lack large-scale reliable relevance annotations.

\lzy{The collections used for metric meta-evaluation should meet three criteria: 1) the size of collections should be large enough to ensure 
the conclusions hold in practical scenarios; 2) the selected collection should contain corresponding human judgements that allow us to align metrics to the gold standard; 3) the dialogues in the collection should be human-to-human to ensure the quality of utterances.}
After a comprehensive survey of existing data sets \cite{cohen2018wikipassageqa, qu2018analyzing, yang2015wikiqa, rajpurkar2016squad, li2017dailydialog, dinan2018wizard, lowe2015ubuntu},
we choose MSDialog \cite{qu2018analyzing} and WizardofWikipedia \cite{dinan2018wizard} collections which meet all three criteria.

MSDialog \cite{qu2018analyzing} is a labeled dialogue dataset of question answering interactions between information seekers and providers from an online forum on Microsoft products. \lzy{This dataset contains more than 2,000 multi-turn information-seeking conversations with 10,000 utterances. Importantly, }there are a variety of human labels in this dataset. In this paper, we use two types of human labels, i.e., `$vote$' and `$is\_answer$', to annotate the relevant answers. \lzy{The left of Table ~\ref{tab:collectionexample} shows an example of MSDialog.  The tag `$vote$' represents the number of `helpful' votes for the answer from the community. If users agree with the response and think this answer may be helpful for this question, users can give one vote to this response. Note that users are not allowed to vote the same response more than once. The tag `$is\_answer$' is a binary tag, which indicates whether this answer is selected as the best answer in the dialogue session. Specially, this tag is often annotated by the user who posted the initial question and started the dialogue. The advantage of this setting is the annotations could be more reliable and closer to the questioner's needs because we can know whether the response really solved the issues.
Therefore, we deem the responses which are labelled as $is\_answer = 1$ to be relevant (i.e., ground truth).}

\begin{table*}[t] 
\centering
\caption{Examples of the selected collections. The responses of `User' and `Agent' are both generated by human beings. }
 \resizebox{\textwidth}{!}{
\begin{tabular}{p{3cm}p{6cm}p{6cm}}
\toprule 
Collection   Name & MSDialog & WizardOfWikipedia \\ \midrule
\multirow{6}{*}{Example} & User: Can anyone   explain what is the Charms menu I haven't a clue. Thanks Kenneth & User: Ive been this girl recently. ... can you give me some advice? \\
 & \textit{\textbf{vote}: 0    \textbf{is\_answer}: 0 }& \textit{\textbf{sentence}: - }\\
 & Agent: The charms menu was something present in Windows 8. You could   move the cursor to the right of the desktop and bring it out. & Agent: Sure, there aren't any hard and fast rules ... it is a   courtship that involves social activities done by the couple either alone or   with others \\
 & \textit{\textbf{vote}: 1    \textbf{is\_answer}: 1} & \textit{}{\textbf{sentence}: It is a form of courtship... }\\
 & User: Thank you & User: Yeah thats true... \\
 & \textit{\textbf{vote}: 0   \textbf{is\_answer}: 0 }& \textit{\textbf{sentence}: - } \\ & & \textit{\textbf{Eval\_score}: 5}\\ \midrule
\multirow{2}{*}{Tag  explanation} & vote: the number of `helpful' votes for the answer from the community. In the original forum, users can annotate each agent answer with a `helpful' vote, which means users agree with the answer and think this answer may be helpful for this question.   & Eval\_score: is the satisfaction score which users are asked to rate over the entire dialogue. The score range   from -1 to 5 (5 is the best) \\
 & is\_answer: whether this answer is selected as the best answer by the community. True is 1 and False is 0. This tag is labelled by the user who asked the question. & sentence: is the raw sentence that agent select from the retrieved sentence pool (`-' means none of the sentences is selected). This sentence come from the original wikipedia document. \\
\bottomrule
\end{tabular}}
\label{tab:collectionexample}
\end{table*}

\begin{figure}
    \centering
    \includegraphics[width=8cm]{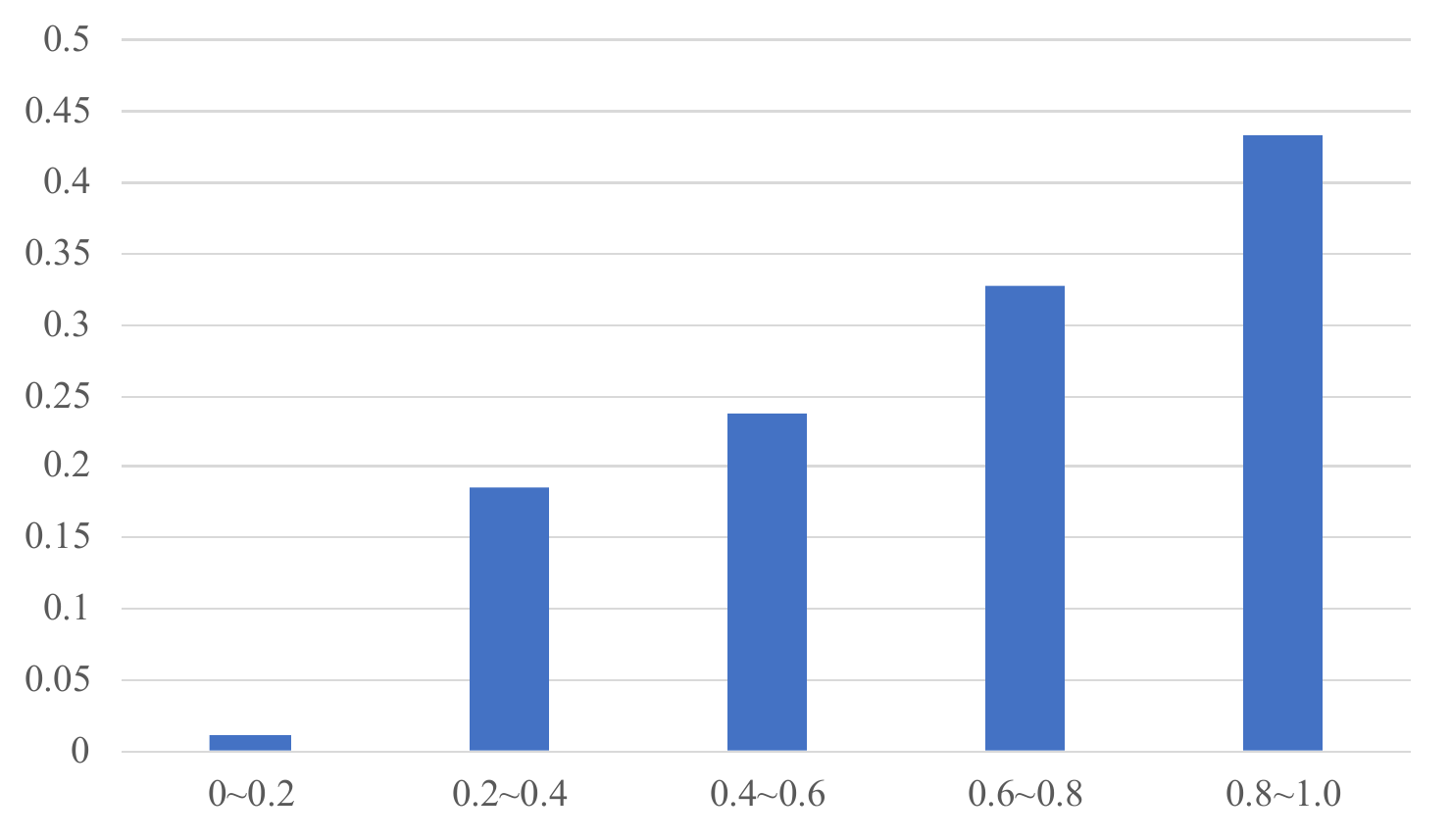}
    \caption{The distribution of ground truth answers in the voted responses. X-axis denotes the normalized vote number, and y-axis denotes the proportion of the ground truth answers.}
    \label{fig:vote_gt}
\end{figure}

\lzy{Moreover, it is necessary to shed light on the connection between `vote' tag and `is\_answer' tag. Here we calculate the proportion of the ground truth answers (i.e., responses with $is\_answer = 1$ tag) in the voted responses. Notice that the exact number of `votes' for each response is not comparable across different conversations because of the number of page views vary across different conversations. For example, trendy questions are more likely to be examined by users and the vote number of their responses are often higher than in low-popularity questions. Therefore, we only focus on the correlation between `vote' and `is\_answer' responses in the same dialogue session. Equation ~\ref{equ:vote} shows the normalization process of the number of votes in each conversation.}
\begin{equation}
    V'_{i,j} = \frac{V_{i,j}}{\max_{k \in j}(V_{k,j}) }  
    \label{equ:vote}
\end{equation}
\lzy{where $V'_{i,j}$ is the normalized vote number of the $i^{th}$-turn response in the $j^{th}$ dialogue, and $V_{i,j}$ is the original vote number of the $i^{th}$-turn response in the $j^{th}$ dialogue. Figure ~\ref{fig:vote_gt}             , which are used in the predictive power test in \S\ref{sec:predictivepower}.}

WizardofWikipedia \cite{dinan2018wizard} is another large supervised dataset with open-domain dialogues grounded with knowledge retrieved from Wikipedia. Unlike MSDialog that is collected from an open source, this collection is built based on a crowd-sourced data collection. In their experiment, the person who answers a question (the wizard) is asked to choose one question-related topic and view a list of sentences (retrieved from Wikipedia) under this topic, and then he/she should select the most relevant sentence and write a response based on the selected sentences. We note that the wizards can also choose nothing and generate the answer using their own words. Therefore, if the wizard chooses one sentence, this means that the sentence is probably relevant to this question. In other words, this sentence selection potentially reveals the relevance between retrieval sentences and questions. \lzy{The right of Table ~\ref{tab:collectionexample} shows an example from this collection. The `Eval\_score' tag is the satisfaction score which users' asked to rate over the entire dialogue. The `sentence' tag is the raw sentence that the agent selects from the retrieved sentence pool. Note this sentence comes from the original wikipedia documents. The value of `sentence' tag can be null if agents do not select any sentences.} Considering all the answers are generated by humans, the answers based on selected sentences are more likely to be relevant to the questions. In the experiment, we regard the answers that have selected sentences 
as ground truth answers.




\subsection{System Runs} \label{sec:systemruns}
\begin{table*}[t] 
\centering
 \resizebox{\textwidth}{!}{
\begin{tabular}{|c|c|c|l|}
\hline
 &  & Type & \multicolumn{1}{c|}{Description} \\ \hline
 & \multicolumn{3}{c|}{Retrieval Model} \\ \hline
1 & BM25 \cite{robertson2004understanding} & \multirow{4}{*}{Classical   Methods} & The   BM25 probabilistic model. \\ \cline{1-2} \cline{4-4} 
2 & TF-IDF \cite{robertson2004understanding} &  & The   tf*idf weighting function \\ \cline{1-2} \cline{4-4} 
3 & LemurTF\_IDF \cite{zhai2001notes} &  & Lemur’s   version of the tf*idf weighting function. \\ \cline{1-2} \cline{4-4} 
4 & Hiemstra\_LM \cite{hiemstra2001using}&  & Hiemstra’s   language model. \\ \hline
5 & BB2 \cite{amati2002probabilistic} & \multirow{9}{*}{DFR framework\footnotemark \cite{harter1975probabilistic}}  & A Bose-Einstein model for randomness \\ \cline{1-2} \cline{4-4}
6 & DFR\_BM25 \cite{amati2002probabilistic} & & The DFR version of BM25 \\ \cline{1-2} \cline{4-4} 
7 & DLH \cite{amati2006frequentist} &  & The DLH hyper-geometric DFR model \\ \cline{1-2} \cline{4-4} 
8 & DPH \cite{amati2006frequentist, amati2008fub}&  & A   different hyper-geometric DFR model using Popper’s normalization \\ \cline{1-2} \cline{4-4} 
9 & IFB2 \cite{amati2002probabilistic}  &  & Inverse Term Frequency model for randomness \\ \cline{1-2} \cline{4-4} 
10 & In\_expB2 \cite{amati2002probabilistic} &  & Inverse   expected document frequency model for randomness \\ \cline{1-2} \cline{4-4} 
11 & InL2 \cite{amati2002probabilistic}&  & Inverse   document frequency model for randomness \\ \cline{1-2} \cline{4-4} 
12 & LGD \cite{clinchant2009bridging, clinchant2010information} &  & A   log-logistic DFR model \\ \cline{1-2} \cline{4-4} 
13 & PL2 \cite{amati2002probabilistic} &  & Poisson   estimation for randomness \\ \hline
 & \multicolumn{3}{c|}{Generative Model} \\ \hline
14 & Seq2seq\_gen \cite{sutskever2014sequence} & \multirow{4}{*}{RNN   framework} & \multirow{2}{*}{A typical RNN-based approach} \\ \cline{1-2}
15 & Seq2Seq\_spec \cite{sutskever2014sequence} &  &  \\ \cline{1-2} \cline{4-4} 
16 & SR\_sum\_gen \cite{tian2017make} &  & \multirow{2}{*}{A hierarchical RNN approach} \\ \cline{1-2}
17 & SR\_sum\_spec \cite{tian2017make} &  &  \\ \hline
18 & Multi-Task\_gen \cite{ghazvininejad2017knowledge, luan2017multi} & \multirow{2}{*}{Multi-task framework} & \multirow{2}{*}{A MemNN-based and Multi-task   learning method} \\ \cline{1-2}
19 & Multi-Task\_spec \cite{ghazvininejad2017knowledge, luan2017multi} &  &  \\ \hline
20 & Transformer\_gen \cite{dinan2018wizard}& \multirow{4}{*}{Transformer   framework} & \multirow{2}{*}{A Generative Transformer Memmory Networks} \\ \cline{1-2}
21 & Transformer\_spec \cite{dinan2018wizard} &  &  \\ \cline{1-2} \cline{4-4} 
22 & TED\_gen \cite{zheng2019enhancing} &  & \multirow{2}{*}{A Transformer with Expanded Decoder Method} \\ \cline{1-2}
23 & TED\_spec \cite{zheng2019enhancing} &  &  \\ \hline
\end{tabular}
}
\caption{The introduction of models used in the experiment. None of these models are specifically conversational search models. In generative models, the suffix \_gen means the models are trained by the third-party collection (dailydialogue \cite{li2017dailydialog}), and suffix \_spec means the models are trained by the subset of the target dataset.}
\label{tab:model_list}
\end{table*}
\footnotetext{Divergence From Randomness (DFR) Framework is introduced in \url{https://github.com/terrier-org/terrier-core/blob/5.x/doc/dfr_description.md} }

\lllzy{System runs play a vital role in the meta-evaluation, for both discriminative power and intuitiveness. Choosing appropriate systems directly affects the reliability of these meta-evaluation approaches. Since modeling conversational search is still a big challenge in IR community \cite{anand2020conversational}, few existing models are specifically designed for conversational search scenarios. In order to cover a broader range of possible cases, we consider both retrieval based models and generative models, which are shown in Table ~\ref{tab:model_list}. Here we deploy 13 different basic retrieval models\footnote{Our work adopts the Terrier IR platform to implement these retrieval models and the details of these models are shown in \url{http://terrier.org/docs/v5.1/configure_retrieval.html} } and 10 different generative models. Note that none of these models are specifically conversational search models, but they are adapted to be used in this context. For example, the retrieval based models simply use the question as the query and then use existing ranking algorithms to rank the candidate responses into the collection as the response. On the other hand, all the generative models we adopted are widely used in dialogue systems, which mainly focus on generating single-turn responses (e.g.,~in the context of question answering) with deep learning models.
Considering the variation of training dataset may affect the performance of generative models, we specially develop generative models with different training settings in the experiment. For example, seq2seq\_gen models are trained by the third-party dataset (DailyDialog \cite{li2017dailydialog}), while seq2seq\_spec models are obtained from the subset of the target dataset (e.g., if the predicting target dataset is MSDialog, we use one subset of MSDialog to train seq2seq\_spec models). Thus, in total, we obtained results from 23 different systems, including 13 retrieval models and 10 generative models, and, in total, have $(23 \times 22)/2=253$ run pairs to compare the performance of the metrics in SRST environments.}

%% file: content/4.0MetaevaluationSingleTurn.tex
In this section, we systematically meta-evaluate the aforementioned metrics (\S\ref{sec:overviewmetric}-\S\ref{sec:multipleturnmetric}) from three perspectives using the data sets (\S\ref{sec:collection}): \emph{reliability} (discriminative power in \S\ref{sec:discriminativepower}), \emph{fidelity} (predictive power in \S \ref{sec:predictivepower}) and \emph{intuitiveness} (\S\ref{sec:intuitiveness}).
\llzy{For the \emph{reliability}, the overall performance of a system is measured by the evaluation metric averaged on all questions and the randomized Tukey's HSD test is then used to determine whether the difference between two runs or systems is statistically significant. For the \emph{fidelity}, two different human-annotated responses (not ground truth) will be selected for each question and these responses are measured by the evaluation metrics. The ratio of the agreement between human choice and the metric's selection (i.e., the response with higher score) is then used to calculate how well the metric correlates with human preference. For the \emph{intuitiveness}, the concordance between gold-standard metrics, which are used to describe different evaluation dimensions, and the evaluation metrics are measured by comparing the agreement between two different runs or systems.}

\subsection{Discriminative Power} \label{sec:discriminativepower}
\input{content/4.1discriminativepower.tex}

\subsection{Predictive Power} \label{sec:predictivepower}
\input{content/4.2predictivepower.tex}

\subsection{Intuitiveness} \label{sec:intuitiveness}
\input{content/4.3intuitiveness.tex}

\subsection{Summary of the Meta-evaluation in the Single-turn Conversational Search} \label{sec:summaryofsingleturn}
\input{content/4.4summaryofsingleturn}

%% file: content/4.1discriminativepower.tex
\lzy{
We introduce the methodology of discriminative power (\S\ref{sec:discriminativemethodology}), the experimental settings (\S\ref{sec:discriminativesettings}), and discuss the results of the SRST and MRST analysis (\S\ref{sec:discriminativeresults}).}

\subsubsection{Methodology} \label{sec:discriminativemethodology}
\ 

\lzy{Discriminative power measures the ability of metrics to detect ``actual'' performance differences which are opposed to the occasional observation \cite{sakai2013metrics, sakai2019diversity}. In other words, discriminative power describes the reliability of metrics to distinguish the differences between systems.} Given a test collection (data set) and a set of runs (conversational search systems), discriminative power is measured by conducting a statistical significance test for every pair of runs and counting the number of significant differences \cite{sakai2012evaluation}. In this paper, we use the randomised version of Tukey's Honestly Significant Differences (HSD) test \cite{carterette2012multiple}. Compared to the bootstrap test~\cite{sakai2006evaluating}, this test is naturally more conservative because the randomised Tukey's HSD test considers the entire set of runs to judge the significance of each run pair. We choose this test because it is suitable for multiple comparisons and reliable on the modern computational power.

\begin{algorithm}[t] 
\caption{Computing Achieved Significance Level \cite{carterette2012multiple}.} 
\label{alg:dpalg1}
\KwIn{A performance value matrix M whose rows represent topics and columns represent runs}
\KwOut{An Achieved Significance Level(ASL) matrix}
\ForEach{pair of runs($r_1$, $r_2$)}{
count($r_1$, $r_2$) \lzyfinal{ = 0}
}
\For{b=1 to B}{
    \For{i=1 to n\tcp*{for each topic(each row of M)}}{
    $i^{th}$ row of $M^{*b}$  =  random permutation of $i^{th}$ row of M\;
    }
    $max^{*b}=max_j\bar{m}^{*b}_j$;$min^{*b}=min_j\bar{m}^{*b}_j$ where \\
    $\bar{m}^{*b}_j$ is the mean of $j^{th}$ column vector of $M^{*b}$ \;
    \ForEach{pair of runs ($r_1$, $r_2$)}{
        \If{$max^{*b}-min^{*b} > |\bar{m}(r_1)-\bar{m}\lzyfinal{(r_2)}|$ where \\
        $\bar{r_j}$ is the mean of column vector for $r_j$ in M}{
        count($r_1$, $r_2$) ++\;
        }
    }
}
\ForEach{pair of runs($r_1$, $r_2$)}{
    ASL($r_1$, $r_2$) = count($r_1$, $r_2$) / B\;
}

\end{algorithm}

The main idea behind Tukey's HSD is that if the largest mean difference observed is not significant, then none of the other differences should be significant either \cite{sakai2013metrics}. Given a set of runs, the null hypothesis is that there is no difference between any of systems. We perform randomised Tukey's HSD as shown in Algorithm ~\ref{alg:dpalg1}, which is taken from \cite{carterette2012multiple}. For a given matrix M whose element (i, j) represents the performance of the $j^{th}$ system for topic (question) $i^{th}$, we create B new matrices $M^b$ by permuting each row random.\footnote{\llzy{Since 
the result of Tukey's HSD can be achieved  when B = 1000 \cite{sakai2012evaluation}, we use this setting in our experiments.}} After that, we compare the performance $\delta$ of each run pair with the largest performance $\delta$ in $M^b$. And then we can obtain the Achieved Significance Level (ASL) for each run pair based on the randomised Tukey's HSD test. Note the null hypothesis is rejected if ASL < $\alpha$, where $\alpha$ is the significance level, and is typically set to 0.05 (95\% confidence level) or 0.01 (99\% confidence level).

\begin{algorithm}[t] 
\caption{Computing the performance $\delta$ \cite{sakai2012evaluation}.} 
\label{alg:dpalg2}
\LinesNumbered

\ForEach{pair of runs($r_1$, $r_2$) with a significant difference at $\alpha$}{
$\delta_\alpha(r_1, r_2) = |mean(r_1)-mean(r_2)|$\;
}
$\delta_\alpha = min_{i,j}\delta_\alpha(r_1, r_2)$
\end{algorithm}

Following previous work \cite{sakai2012evaluation, zhou2013reliability}, we also estimate the performance $\delta$ required to achieve a statistical significance at $\alpha$ for a given topic set size (shown in Algorithm ~\ref{alg:dpalg2}). We take the smallest $\delta$ from all the run pairs that are significantly different. 

\subsubsection{Experiment settings} \label{sec:discriminativesettings}
\ 

Different strategies are adopted to generate the responses in different scenarios. In the SR case, \llzy{for retrieval models,} the top 1 result of the search result list is selected as the proposed response. This case is similar to dialogue systems where one response is for one question. 

In the MR case, \llzy{since generative models are rarely used to produce response lists, we only consider retrieval-based models (i.e., Model 1\textasciitilde 13 in Table ~\ref{tab:model_list}) in this case. Therefore, in total, we have $(13 \times 12 )/ 2 = 78$ run pairs in MRST. }\lzy{Although the layout of the responses is similar to the traditional ranking list, there are still some considerable differences between these list-like responses and traditional result lists: 
(1) The number of search results in the responses is smaller than that in the traditional web search. Limited by the design of the interactive user interface (e.g., screen size), conversational search systems usually provide only a few search results in one response. (2) Users may regard the result list as a whole response, especially when the list contains the expected information need. 
}
Given those unique differences, we extract the top 5 retrieved responses from the search result list. 
To evaluate those with ranking-based metrics, we exploit the ranking order within the original search result list for such evaluation.
On the other hand, when evaluating based on word based metrics (i.e., BLEU, METEOR , ROUGE, EA, SCS, BERTScore and BERT-RUBER), we concatenate these retrieved results, to reflect the scenario that users view the entire ranked list as one response. 

\subsubsection{Experiment results} \label{sec:discriminativeresults}
\



\begin{figure}
    \includegraphics[width=\textwidth]{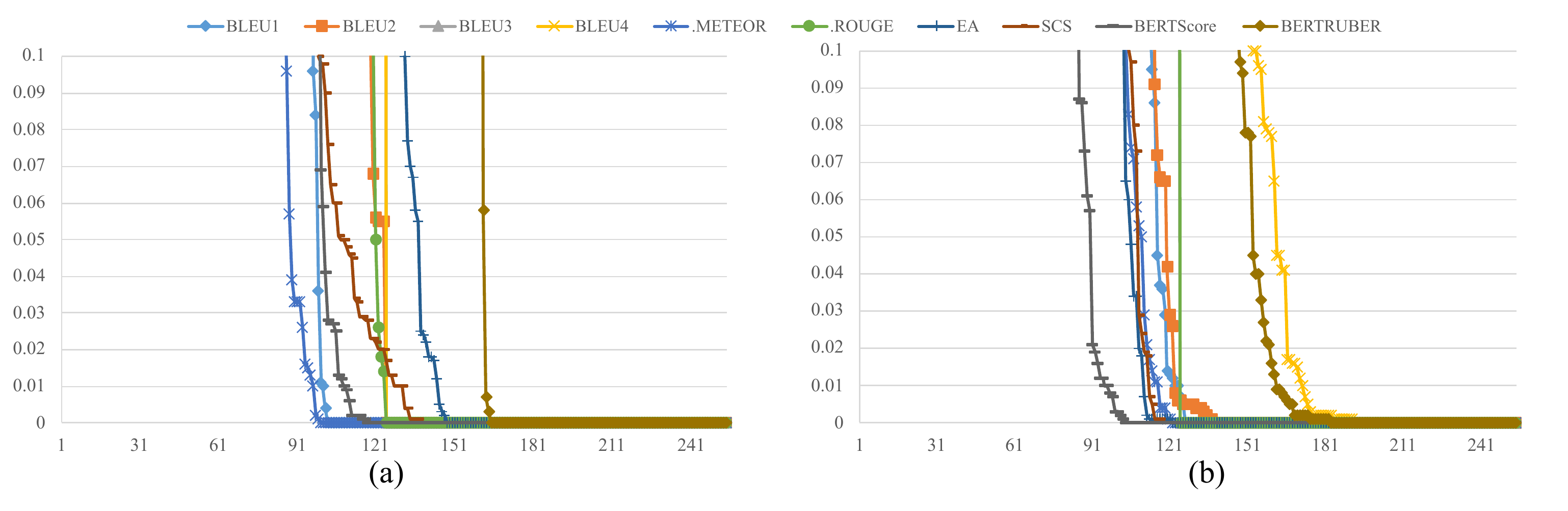}
    \caption{Discriminative power evaluation in SRST scenario for two collections: (a) MSDialog and (b) Wizardofwikipedia. ASL curves are calculated based on the randomised Tukey's HSD. y-axis:ASL (i.e., p-value); x-axis: run pairs sorted by ASL.}
    \label{fig:asl_single_turn}
\end{figure}

\noindent \textbf{(a) SRST Scenario:} \lzy{In this section, we evaluate all the SR metrics listed in the second column of Table ~\ref{tab:metric}, including word overlap-based, embedding-based metrics and learning-based metrics.}
Figure ~\ref{fig:asl_single_turn} shows the ASL curves of those selected conversational search metrics, based on the Randomized Tukey's HSD on MSDialog and WizardofWikipedia data sets. 
The metrics that are closer to the origin (left) are more discriminative.

Table  ~\ref{tab:discpower} cuts Figures ~\ref{fig:asl_single_turn} in half vertically at \lzyfinal{$\alpha = 0.05$} to quantify discriminative power and the performance $\delta$ required for achieving statistical significance with a given number of conversational search sessions (randomly sampling 1000 sessions from MSDialog and WizardofWikipedia, respectively). \llzy{For example, the left side of Table ~\ref{tab:discpower} shows that on the MSDialog data set, the discriminative power of BLEU2 according to the Tukey's HSD test at $\alpha = 0.05$ is 130/253 = 51.4\% (i.e.,~130 significantly different run pairs were found) and $\delta$ required for achieving achieving statistical significance is 0.06. The higher the percentage of significantly different run pairs, the more discriminative the metric maintains.}


By comparing the different metrics in terms of discriminative power in different 
data sets as shown in Figure ~\ref{fig:asl_single_turn} and Table ~\ref{tab:discpower}, we summarise our findings below:

\llzy{
\begin{itemize}
    \item Generally, it can be observed that both word overlap-based and embedding-based metrics perform better than the learning-based metric, which means the fine-tuned learning-based metric may not be good at discriminating different models in SRST. Therefore, we do not recommend the choice of learning-based metrics to compare the difference between models in the single-turn conversational search.
    \item The performance of word overlap-based metrics and embedding-based metrics may vary in different datasets. We found that word overlap-based metrics (Avg.=55.7 on disc. power) are generally better than embedding-based metrics (Avg.=54.67) in MSDialog, while embedding-based metrics (Avg.=64.5) outperform word overlap-based metrics (Avg.=50.8) in WizardofWikipedia. From this observation, it is better to consider both types of metrics if we compare different models using different collections.
    \item METEOR achieved the best performance amongst word overlap-based metrics, and BERTScore outperforms other embedding-based metrics across the two selected datasets. Moreover, we can see the discriminative performance of METEOR is very close to the performance of embedding-based metrics. Especially when the embedding-based metrics generally achieve higher discriminative scores than word overlap-based metrics in WizardofWikipedia collection, the performance of METEOR is still similar to SCS. Compared to embedding-based metrics, METEOR can be more interpretable.
    \item In the BLEU metric set, it is worth noting that BLEU1 is more discriminative than other BLEU metrics (i.e., BLEU2, BLEU3 and BLEU4) for both collections. This might be because the ratios of overlap n-gram ($n \geq 2$) between responses and references are low. It is difficult for n-gram BLEU to detect the difference when both responses have a low proportion of word overlapping. 
\end{itemize}
}

\begin{table}[h]
\caption{Discriminative power of metrics based on the randomised Tukey's HSD test at 0.05 on the MSDialog and WizardofWikipedia collections in SRST scenario. \llzy{$\delta$ is the threshold that required for achieving statistical significance.}}
\begin{tabular}{|c|c|c|c|c|}
\hline
\multirow{2}{*}{Metric} & \multicolumn{2}{c|}{MSDialog} & \multicolumn{2}{c|}{WizardofWikipedia} \\ \cline{2-5} 
 & Disc. Power & $\delta$ & Disc. Power & $\delta$ \\ \hline
BLEU1 & 61.7 & 0.04 & 54.9 & 0.03 \\
BLEU2 & 51.4 & 0.06 & 53.4 & 0.02 \\
BLEU3 & 51.4 & 0.07 & 51.4 & 0.03 \\
BLEU4 & 51.4 & 0.06 & 36.8 & 0.02 \\
METEOR & \textbf{65.6} & 0.04 & \textbf{56.9} & 0.03 \\
ROUGE & 52.6 & 0.04 & 51.4 & 0.10 \\ \hline
EA & 46.2 & 0.02 & 58.9 & 0.02 \\
SCS & 57.3 & 0.05 & 57.7 & 0.04 \\
BERTScore & \textbf{60.5} & 0.01 & \textbf{64.8} & 0.01 \\ \hline
BERT-RUBER & 36.4 & 0.08 & 40.3 & 0.06 \\\hline
\end{tabular}
\label{tab:discpower}
\end{table}

From the above findings, we can see that high discriminative power comes mostly from the embedding-based metrics. Interestingly, METEOR, which is based on word overlap, was also observed to perform well in discriminating systems. That might demonstrate that the metrics which take into account the similarity/connection between words might be more likely to be statistically reliable and consistent in conversational search scenarios. 

To take a step further, Table ~\ref{tab:discpoweranalysis} provides a detailed analysis on the discriminative power results of selected metrics from Table ~\ref{tab:discpower}. \lzy{This table presents the distribution of significant run pairs, which are discriminated by the metrics, and further show the degree of overlap between the significantly different run pairs for all the selected metrics.} 
\llzy{The comparison results are presented as X/Y/Z form. X indicates the number of pairs that are significant for the row metrics but not for the column ones, and Z shows the number of pairs that are significant for the column metrics but not for the row ones. Y means \lzyfinal{the number of significant pairs that are found by both metrics}. For example, in MSDialog, it can be seen that BLEU1 and 
BLEU2 have 129 run pairs in common, and that these two metrics obtained 27+129=156 significant differences and 1+129=130 significant differences, respectively (these correspond to the discriminative power values of 156/253=61.7\% and 130/253=51.4\%). A large Y means that the two metrics may be more similar to each other, and the larger X (or Z) indicates that the row (or column) metrics are more discriminative than the column (or row) ones. Note that the case where two metrics give the opposite results that are both significant might occur if the significant runs, which are detected by these two metrics, are totally different from each other. This means that metrics with similar discriminative power scores may also present different discriminative abilities on different run pairs.} \lllzy{However, this is rare and we do not observe this in our experiment.} 
From this table, it can be observed that:


\begin{itemize}
    \item METEOR is quite similar to embedding-based metrics in both collections. For example, we can find that METEOR and EA have 110 run pairs in common for MSDialog and 128 run pairs for WizardofWikipedia. At the same time, METEOR and SCS have 130 and 134 common pairs in MSDialog and WizardofWikipedia, respectively. These overlap degrees are much higher than other word overlap-based metric pairs. 
    \item \llzy{The performance of BLEU metrics with different settings are very similar. From Table ~\ref{tab:discpoweranalysis}, it can be observed that BLEU2, BLEU3 and BLEU4 share high percentage of common pairs with each other, especially in MSDialog. BLEU2-4 may have similar ability in discriminating systems.}
\end{itemize}

\begin{table}[t] 
\caption{Comparison of significantly different run pairs on the MSDialog and WizardofWikipedia (randomised Tukey's HSD at $\alpha = 0.05$). In each table cell, the number of significant run pairs for metrics are presented as X/Y/Z. X is the number of run pairs that are significant for the row metrics but not for the column ones, and Z shows the number of pairs that are significant for the column metrics but not for the row ones. \lzyfinal{ Y indicates the number of significant pairs that are found by both metrics.} }
 \resizebox{\textwidth}{!}{
\begin{tabular}{|c|ccccccccc|}
\hline
\multicolumn{10}{|c|}{MSDialog} \\ \hline
 & BLEU2 & BLEU3 & BLEU4 & METEOR & ROUGE & EA & SCS & BERTScore & BERTRUBER \\ \hline
BLEU1 & 27/129/1 & 27/129/1 & 27/129/1 & 0/156/10 & 25/131/2 & 51/105/12 & 33/123/22 & 15/141/12 & 66/90/2 \\
BLEU2 &  & 0/130/0 & 0/130/0 & 0/130/36 & 0/130/3 & 52/78/39 & 34/96/49 & 13/117/36 & 65/65/27 \\
BLEU3 &  &  & 0/130/0 & 0/130/36 & 0/130/3 & 52/78/39 & 34/96/49 & 13/117/36 & 65/65/27 \\
BLEU4 &  &  &  & 0/130/36 & 0/130/3 & 52/78/39 & 34/96/49 & 13/117/36 & 65/65/27 \\
METEOR &  &  &  &  & 33/133/0 & 56/110/7 & 36/130/15 & 23/143/10 & 76/90/2 \\
ROUGE &  &  &  &  &  & 52/81/36 & 34/99/46 & 16/117/36 & 68/65/27 \\
EA &  &  &  &  &  &  & 3/114/31 & 24/93/60 & 26/91/1 \\
SCS &  &  &  &  &  &  &  & 28/117/36 & 55/90/2 \\
BERTScore &  &  &  &  &  &  &  &  & 64/89/3 \\ \hline
\multicolumn{10}{|c|}{WizardOfWikipedia} \\ \hline
 & BLEU2 & BLEU3 & BLEU4 & METEOR & ROUGE & EA & SCS & BERTScore & BERTRUBER \\ \hline
BLEU1 & 19/120/15 & 29/110/20 & 46/93/0 & 5/134/10 & 29/110/20 & 17/122/27 & 12/127/19 & 10/129/35 & 52/87/15 \\
BLEU2 &  & 10/125/5 & 42/93/0 & 15/120/24 & 10/125/5 & 27/108/41 & 25/110/36 & 9/126/38 & 56/79/23 \\
BLEU3 &  &  & 37/93/0 & 20/110/34 & 0/130/0 & 32/98/51 & 30/100/46 & 9/121/43 & 60/70/32 \\
BLEU4 &  &  &  & 0/93/51 & 0/93/37 & 1/92/57 & 1/92/54 & 7/86/78 & 24/69/33 \\
METEOR &  &  &  &  & 34/110/20 & 16/128/21 & 10/134/12 & 12/132/32 & 50/94/8 \\
ROUGE &  &  &  &  &  & 32/98/51 & 30/100/46 & 9/121/43 & 60/70/32 \\
EA &  &  &  &  &  &  & 10/139/7 & 15/134/30 & 49/100/2 \\
SCS &  &  &  &  &  &  &  & 15/131/33 & 47/99/3 \\
BERTScore &  &  &  &  &  &  &  &  & 75/89/13 \\ \hline
\end{tabular}}
\label{tab:discpoweranalysis}
\end{table}

\noindent \textbf{(b) MRST Scenario.}
%
Next we discuss the performance of the metrics in the MRST scenario. As shown in Figure ~\ref{fig:asl_multiple_turn} and Table ~\ref{tab:discpower_mr}, we compare the different word-based metrics and modified ranking-based metrics in terms of their discriminative power. 
\llzy{In the experiments, we noticed that the performance of ranking-based metrics (e.g. nDCG) have a similar trend with the metrics which are used to estimate the information gain in Equation ~\ref{equ:gain}. In other words, if the better SR metrics are used in the gain calculation, the ranking-based metrics could achieve better performance. For example, when METEOR > BLEU2 > ROUGE in SRST (shown in Table ~\ref{tab:discpower}), it can be observed that nDCG(METEOR) > nDCG(BLEU2) > nDCG(ROUGE) in Table ~\ref{tab:discpower_mr}. Since it is a compromised way to use direct SR metrics to estimate the information gain, using such metrics is not a key concern when we compare the retrieval-based metrics. Therefore, to reduce the complexity of our experiment, we select METEOR, which achieves the best performance in SRST evaluation among the word overlap-based metrics, to compare the ranking-based metrics. }Our findings are summarised as below:

\begin{figure}
    \includegraphics[width=\textwidth]{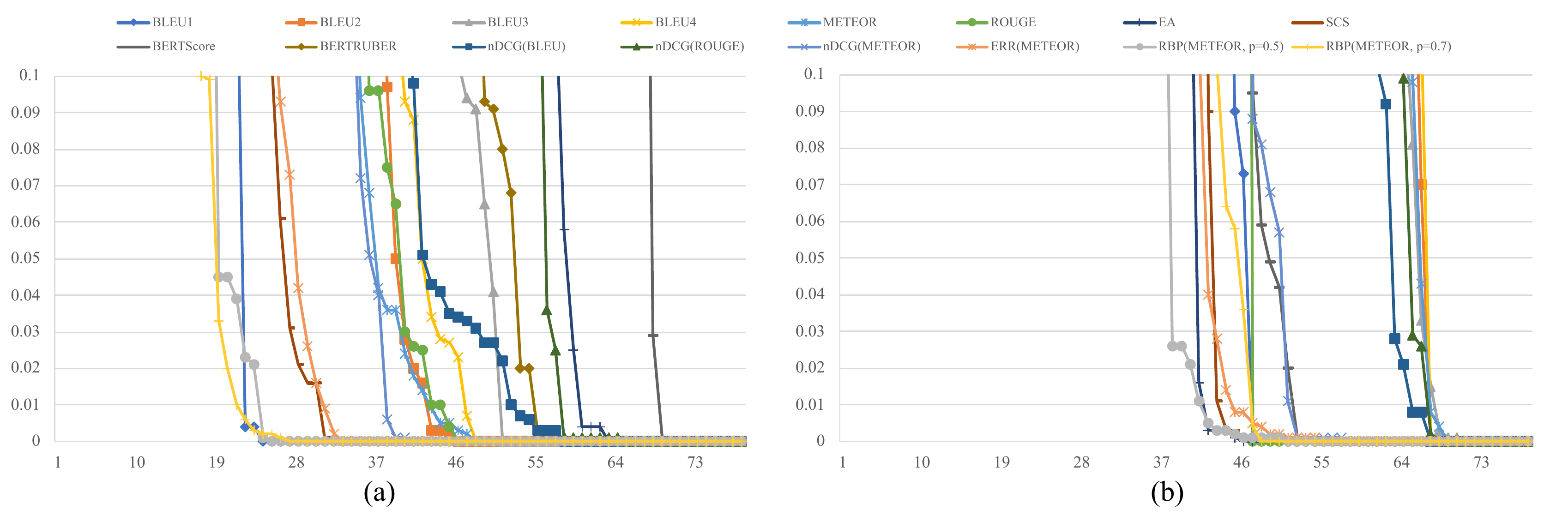}
    \caption{Discriminative Power Evaluation in MRST scenario: ASL curves based on the randomised Tukey's HSD. y-axis:ASL (i.e., p-value); x-axis: run pairs sorted by ASL. (a) is results of MSDialog collection and (b) is results of Wizardofwikipedia collection.}
    \label{fig:asl_multiple_turn}
\end{figure}

\begin{table}[h] 
\caption{Discriminative power of MR metrics based on the randomised Tukey's HSD test at 0.05 on the MSDialog and WizardofWikipedia collections in MRST scenario. \llzy{$\delta$ is the threshold that required for achieving statistical significance. N/A means the metric detect no significant pairs in the corresponding collection.}}
\begin{tabular}{|c|c|c|c|c|}
\hline
\multirow{2}{*}{Metric Name} & \multicolumn{2}{c|}{MSDialog} & \multicolumn{2}{c|}{WizardofWikipedia} \\ \cline{2-5} 
 & Disc. Power & $\delta$ & Disc. Power & $\delta$ \\ \hline
BLEU1 & 73.1 & 0.01 & 41.0 & 0.01 \\
BLEU2 & 50.0 & 0.01 & 15.4 & 0.01 \\
BLEU3 & 37.2 & 0.01 & 16.7 & 0.01 \\
BLEU4 & 46.2 & 0.01 & 15.4 & 0.01 \\
METEOR & 53.8 & 0.01 & 16.7 & 0.01 \\
ROUGE & 50.0 & 0.01 & 41.0 & 0.01 \\ \hline
EA & 25.6 & 0.01 & 48.7 & 0.01 \\
SCS & 66.7 & 0.01 & 46.2 & 0.01 \\ 
BERTScore & 14.1 & 0.02 & 38.5 & 0.01 \\ \hline
BERT-RUBER & 33.3 & 0.03 & 0.0 & N/A \\ \hline
nDCG(BLEU2) & 46.2 & 0.03 & 20.5 & 0.03 \\
nDCG(ROUGE) & 29.5 & 0.02 & 17.9 & 0.02 \\ \hline
nDCG(METEOR) & 53.8 & 0.02 & 35.9 & 0.02 \\
RBP(METEOR, p=0.5) & 76.9 & 0.01 & 52.6 & 0.01 \\
RBP(METEOR, p=0.7) & 76.9 & 0.01 & 42.3 & 0.01 \\
ERR(METEOR) & 65.4 & 0.01 & 47.4 & 0.01 \\ \hline
\end{tabular}
\label{tab:discpower_mr}
\end{table}

\begin{itemize}
    \item Modified ranking-based metrics with METEOR are generally more discriminative than most of word overlap-based metrics. It is worth noting that RBP with METEOR is the most discriminative metric in both MSDialog and WizardofWikipedia. Specially, it can be observed that RBP can maintain a high level of performance when \emph{$p=0.5$} although the results on the two collections vary.
    \item In the word overlap-based metrics, BLEU1 performs comparatively well in both 
    collections. 
    \item Compared with the results of the SRST scenario, we find the discriminative performance of metrics may vary significantly in different scenarios. For example, \llzy{BERTScore performs poorly in MRST settings, while it outperforms other metrics in SRST. On the contrary, while ROUGE performs at the average level in SRST, it proves to be much more powerful in MRST scenarios in terms of discriminative power, especially in WizardofWikipedia collection.} \lzy{This is because all the candidate responses are longer than the ground truth, which increases the variation of response recall. As a recall-oriented metric, ROUGE might be more sensitive to this variation.}
    \item \llzy{It is also worthwhile noting that the learning-based metrics still perform poorly in MRST scenarios. Since the unreferenced models could be varied by the dataset, it is observed that the performance of BERT-RUBER is not robust in different datasets. For example, we can see BERT-RUBER cannot detect any significant pairs in the WizardofWikipedia collection.}
\end{itemize}

%% file: content/4.2predictivepower.tex
We describe the methodology of predictive power (\S\ref{sec:predictivemethodology}), followed by the experiment setting (\S\ref{sec:predictivesettings}). Since both two collections do not have human annotations on multiple responses, only the SRST results are presented (\S\ref{sec:predictiveresults}).

\subsubsection{Methodology} \label{sec:predictivemethodology}
\

\begin{algorithm}[t] 
\caption{Computing predictive power.} 
\label{alg:pdalg}

Total=0;Correct=0\;
\For{d = 1 to N \tcp{for each dialogue}}{
\ForEach{pair of responses($r_1$, $r_2$) in dialogue d}{
Total++\;
$\delta X=X(r_1)-X(r_2)$\; \tcp{$X$ is the metric score of response $r_j$}
$\delta X^*=X^*(r_1)-X^*(r_2)$\; \tcp{$X^*$ is the human judgement score of response $r_j$, such as votes}
\If{($(\delta X \times \delta X^*) > 0 $) \tcp{$X$ and $X^*$ positively agree}}{
    Correct++\;
}
}
}
PredictivePower = Correct/Total\;

\end{algorithm}

Predictive power measures the ability of metrics to \lzy{describe the agreement between metrics and user preferences~\cite{sanderson2010user}. In other words, predictive power reflects the \emph{fidelity} of the metrics of measuring ultimate user preferences.}
To evaluate evaluation metrics, a natural way is to ask real users to check whether evaluation metrics are consistent with their judgements. However, it is difficult to involve real users in determining which metric is better in practice. Therefore, the basic idea of predictive power is that if an evaluation metric agrees with the user's preference between two outputs, then that is a correct prediction \cite{sakai2013metrics}. 
The higher the predictive power score is, the better or more similar to the human judgements the metric is. We use predictive power to examine the similarity between metrics and human judgements in conversational search (as shown in Algorithm ~\ref{alg:pdalg}). 

\subsubsection{Experiment settings} \label{sec:predictivesettings}
\

With the definition of the predictive power, we should first generate the collections that consist of response pairs (X, Y) with different human judgements. 
\lzy{For MSDialog, }we first extract all the possible combination pairs from each conversation. 
In each pair, the response with more `votes' is selected as the `preferred' answer. 
Note we discard the pair if the votes of the two responses are equal.
\lllzy{Since the ratio of `is\_answer' in the responses with less than five votes is very low (5.92 \%), most of these responses may not be able to answer the questions. } To reduce the effect of low quality responses, only the responses with more than five `votes' are selected as the `preferred' answers. 

\lzy{As described in section \ref{sec:collection}, different from MSDialog, WizardofWikipedia only has binary annotations in sentence selection, which means we only know which sentence is selected by the agents and which are not. In order to generate the user preference pairs for this corpus, all the sentences under the selected topics are extracted and grouped randomly into pairs. The selected sentences are regarded as `preferred'. The sentence pairs can be discarded if both of the sentences in the pair are not selected by the agent. After those steps, we have the user preference data for both data sets for meta-evaluating the predictive power of metrics.} 


\subsubsection{Experiment results} \label{sec:predictiveresults}
\ 


Table ~\ref{tab:predpower} presents the predictive score for all the SR metrics (shown in Table ~\ref{tab:metric}), including word overlap-based (BLEU, METEOR and ROUGE), embedding-based (EA, SCS and BERTScore) and learning-based metrics (BERT-RUBER). The scores in the table indicate the agreements between metrics and user preferences. For example, the left side of Table ~\ref{tab:predpower}(a) shows metric BLEU1 has 54.16\% `correct' prediction, which agrees with users' preference (i.e., selecting higher `votes' responses) within all the sessions in the MSDialog collection. Note, due to the completely different generation process of these two collections (see \S\ref{sec:collection}), it is unfair to compare the numerical values between these two collections. For example, it is observed that the predictive power of BLEU1 is 54.16\% in MSDialog and 91.16\% in WizardofWikipedia, which says nothing about which collection can allow this metric to perform better because the settings of these two collection are very different. 
Moreover, in order to evaluate the performance of these metrics, we also present the result of random selection (i.e., randomly selecting one answer as the `preferred' answer). \lllzy{We use two-tailed T-test significant test to examine the difference between metrics and baseline. } By comparing the different metrics in terms of predictive power as shown in Table ~\ref{tab:predpower}, it can be observed that:

\begin{itemize}
    \item \llzy{The learning-based metrics appear to be more correlated to the human preference. It is observed that BERT-RUBER comparatively outperforms other metrics in terms of predictive power in MSDialog. However, the performance of learning-based metrics are not robust in different collections. From Table ~\ref{tab:predpower}, we can find BERT-RUBER performs poorly in the WizardofWikipedia collection due to the influence of its training dataset. This may limit the application of learning-based metrics.}
    \item \llzy{Except for BERT-RUBER, }METEOR is consistently better than other word overlap-based metrics in both MSDialog and WizardofWikipedia collection. However, we observe that the difference between the metrics is very small (around 1\% to 2\%).
    \item Word embedding-based metrics are generally less predictive than word overlap-based metrics in the WizardofWikipedia collection. Unlike the outstanding performance of discriminative power, it is observed that embedding-based metrics perform poorly in predicting the judgements. In the WizardofWikipedia collection, EA is the least predictive metric for evaluating conversational search. This is because word embedding-based metrics may consider more similar words than overlap-based metrics, which may incorporate a lot of noise and thus affect the predictive performance of the metrics.
    \item It is worth noting that 
    \lllzy{the predictive power of all the metrics outperform the random baseline, but are still much worse than the human judgment approach (with a predictive power of 1.0) in MSDialog. This difference, to some extent, also shows the weak correlation between these metrics and human judgement, which is consistent with the conclusions of prior work \cite{liu2016not, novikova2017we}. }
    After a close examination, we find that the metric score distribution of `correct' answers (i.e., the answer which have more votes) and `wrong' answers are very similar. For example, for metric BLEU4, the average score of a `correct' answer is 0.006, while the average score of a `wrong' answer is 0.003. The median of these two kinds of answers are also very close. This means that both `correct' and `wrong' answers may not have good word overlapping.  \lllzy{Moreover, we can see the predictive power of most metrics in WizardofWikipedia reach more than 90\%. This is because the word overlapping rate between selected sentences and ground truth is very high, while the unselected sentences always have few common words with the ground truth. This huge difference between selected sentences and unselected sentence may overestimate the performance of word overlap-based metrics. Therefore, although the predictive power of metrics in WizardofWikipedia are near 100\%, it is still inadequate to prove these metrics have good correlation with human judgements.}
    \item We find that all the predictive power measures in the WizardofWikipedia collection are more than 90\%. This is not surprising since the users may prefer to write the answers based on the selected sentences, and the `correct' answers may have more words overlapping with the sentences that they choose.  
\end{itemize}

The above observations suggest that METEOR may be a good metric to predict users' preferences in conversational search scenarios (top 3 metric in both collections) \llzy{if we consider accuracy and robustness of metrics}. However, \llzy{except for BERT-RUBER,} the correlation between metrics' prediction and human judgements is very weak \llzy{because we can find all the metrics have minor improvements over baseline (random selection strategy) in MSDialog and little variation among different metrics in terms of predictive power.}

\begin{table}[] 
\caption{Predictive power of metrics on the MSDialog and WizardofWikipedia collections. \llzy{Statistical significance is determined by two-tailed t-test with Bonferroni correction, where a value of p < 0.01/10 = 0.001 is annotated as \textbf{**} and p < 0.05/10 = 0.005 is annotated as \textbf{*}}}
\begin{tabular}{ccc}
\toprule
Metric Name & MSDialog & WizardofWikipedia \\ \midrule
random(baseline) & 50.02\% & 50.18\% \\ \midrule
BLEU1 & 54.16\%** & 91.16\%** \\
BLEU2 & 53.50\%** & 91.99\%** \\
BLEU3 & 52.04\%** & 91.53\%** \\
BLEU4 & 53.37\%** & 91.23\%** \\
METEOR & 55.65\%** & \textbf{93.98}\%** \\
ROUGE & 54.12\%** & 92.46\%** \\ \midrule
EA & 53.72\%** & 79.05\%** \\
SCS & 53.15\%** & 89.37\%** \\
BERTScore & 56.84\%** & 90.76\%** \\ \midrule
BERT-RUBER & \lzyfinal{\textbf{72.95}\%**} & 68.59\%** \\ \bottomrule
\end{tabular}
\label{tab:predpower}
\end{table}

%% file: content/4.3intuitiveness.tex
We introduce the method of testing intuitiveness and our settings in \S\ref{sec:intuitivenessmethodology} and \S\ref{sec:intuitivenesssettings}. 
We discuss the intuitiveness results (\S\ref{sec:intuitivenessresults}) for SRST only, given the difficulty of applying intuitiveness test for the MRST scenario without suitable gold standard metrics (see \S\ref{sec:intuitivenessmethodology}).

\subsubsection{Methodology} \label{sec:intuitivenessmethodology}
\ 

This section discusses the method of a concordance test that examines the intuitiveness of metrics. A concordance test is a user-free version of the predictive power test \cite{sakai2012evaluation} and has been widely applied to measuring intuitiveness for diversity IR metrics \cite{zhou2013reliability, chen2017meta}. 
Different from predictive power that examines the ultimate user preference, the aim of this method is to quantify `which metric is more intuitive' and examine how `intuitive' it is. 
\lllzy{With the concordance test, we can further dig into the specific dimensions of preference we are interested in, rather than be restricted to the overall user preference. 
Algorithm ~\ref{alg:ctalg} shows the concordance test algorithm for comparing two candidate metrics M1 and M2, given a gold standard metric M*. The gold standard is always an automatic metric and represents a basic property that the candidate metric should satisfy. 
}
We conduct the concordance test in the following steps: firstly, all pairs of systems are collected for testing the disagreements between M1 and M2. 
Then, out of these disagreements, we calculate how often each metric agrees with the gold standard metric. By comparing the agreements between the two metrics, we can analyze which metric focuses more on one specific dimension \footnote{In intuitiveness, we follow the previous studies \cite{chen2017meta, sakai2013metrics, sakai2012evaluation} and only consider two metrics' disagreements or agreements with a gold metric. When two metrics agree, they are ignored. However, if there are no disagreements between the two metrics, then the concordance score of the two metrics are both equal to 0, which further make both metrics disagree with gold standards. In this case, it may be not sufficient to understand which metric is the closest to the gold standard.}.

\begin{algorithm}[t] 
\caption{Computing the concordance of M1 and M2.}
\label{alg:ctalg}

Disagreements = 0; $Conc_1$ = 0; $Conc_2$ = 0\;
\ForEach{pair of runs($r_1$, $r_2$)}{
\ForEach{topic t}{
$\delta M1 = M1(t, r_1) - M1(t, r_2)$\;
$\delta M2 = M2(t, r_1) - M2(t, r_2)$\;
$\delta M^* = M^*(t, r_1) - M^*(t, r_2)$\;

\If{($\delta M1 \times \delta M2 < 0$)\tcp{M1 and M2 disagree}}{
    Disagreements ++\;
    \If{($\delta M1 \times \delta M^* \geq 0$)\tcp{M1 and $M^*$ agree}}{$Conc_1$ ++\;}
    \If{($\delta M2 \times \delta M^* \geq 0$)\tcp{M2 and $M^*$ agree}}{$Conc_2$ ++\;}
}
}
}
$Conc(M1 | M2, M*) = Conc_1/Disagreements$\;
$Conc(M2 | M1, M*) = Conc_2/Disagreements$\;
\end{algorithm}

For evaluation of conversational search (or any dialogue models), many dimensions were proposed to measure the performance of the responses (conversations), such as adequacy \cite{papineni2002bleu}, fluency \cite{liu2016not}, informativeness \cite{liu2016not} and human likeness \cite{deriu2020survey}. In this paper, we mainly focus on three key aspects that were widely adopted, namely, \emph{adequacy}, \emph{fluency} and \emph{informativeness}. Since it is impractical to perform human annotations for the two large-scale collections we use, we choose several `gold standard' metrics to substitute for manual annotations following 
\citet{sakai2013metrics}.

We review existing literature to find appropriate `gold standard' metrics for each of these three aspects:

\begin{itemize}
    \item \textbf{Adequacy} is a measurement that describes the presence of correct meaning in the answers, 
    which indicates mainly on 
    the relevance of answers' content to the question. Some studies have discussed the correlation between human annotations and conversation metrics in terms of adequacy. For example, \citet{liu2016not} compare word overlap-based and word embedding-based metrics with human judgment in both Twitter and Ubuntu collections. 
    They found BLEU2, a reference-based metric, 
    has the highest correlation with the adequacy annotations. 
    \llzy{It is worthwhile noting that some previous studies propose non-reference-based metrics to estimate the adequacy in the dialogue evaluation, such as \cite{sinha2020learning}. However, these metrics are learning-based methods and might also be influenced by the training settings. It is difficult to interpret when using these metrics as gold standard metrics. Although BLEU2 is a reference-based metric, it is sufficiently simple to interpret and has been empirically shown to outperform others in capturing adequacy.}
    Therefore, we choose BLEU2 as the gold standard for adequacy.
    
    \item \textbf{Fluency} is a metric that reflects how well-formed the answers are in a language. 
    SLOR is one of the state-of-the-art metrics for reference-less fluency evaluation and has a good correlation with the human judgments at the sentence level \cite{kann2018sentence}. 
    SLOR calculates the sentence score based on the log-probability under a given language model and unigram log-probability as follows:

\begin{equation}
    SLOR(S) = \frac{1}{|S|}(ln(p_{LM}(S)))-ln(p_{u}(S)))
\end{equation}

where $p_{LM}(S)$ is the probability of the sentence $S$ under the language model, $|S|$ refer to the length of the sentence. In our study, we apply a trigram language model to calculate the $p_{LM}(S)$. The unigram probability $p_{u}(S)$ is calculated as

\begin{equation}
    p_{u}(S)=\prod_{t \in S} p(t)
\end{equation}

where p(t) is the unconditional probability of one token t.

    \item \textbf{Informativeness} reveals the ability of candidate answers to provide useful information for the questions. \lzy{
    Different from adequacy, an answer might be adequate (relevant) to the question, but not informative (i.e.,~not providing any useful or new information). 
    Similar to fluency, informativeness also lacks available automatic measurements in the dialogue evaluation field.}
    Studies often ask humans to rate informativeness directly\cite{wong2011comparative}. 
    However, Novikova et al. \cite{novikova2017we} performed an experiment to compare the word-based metrics (e.g., BLEU) and grammar-based metrics with human annotations in terms of informativeness. They found LEPOR metric \cite{han2012lepor} achieved the highest correlation scores across three different corpora in terms of informativeness. 
    LEPOR is a language independent machine translation metric, which consists of sentence length penalty, n-gram position difference penalty, precision, and recall. 
    \lllzy{Since this is the best informativeness metric empirically found to be aligning with human annotation, we choose LEPOR as the gold standard for informativeness \cite{novikova2017we}.}
    It is calculated as: 
    
    \begin{equation}
        LEPOR = LP \times NPosPenal \times Harmonic(\alpha Rec, \beta Prec)
        \label{equ:lepor_1}
    \end{equation}
    
    where LP means Length Penalty, which is defined to consider the penalty for both longer and shorter system outputs compared with the reference. If the candidate length of sentence is larger or smaller than that of the reference, LP will be less than one which gives a penalty on the evaluation value of LEPOR. NPosPenal aims to compare the words order in the sentences between reference and candidate. It is defined as:
    
    \begin{equation}
        NPosPenal=e^{-NPD}
    \end{equation}
    
    \begin{equation}
        NPD = \frac{1}{S_{cand}}\sum_{S_{cand}}^{i=1} \left | PD_i \right |
    \end{equation}
    
    where $S_{cand}$ is the length of candidate sentence and $PD_i$ represents the n-gram position difference value of aligned words between candidate and reference sentences. The parameter $Harmonic(\alpha Rec, \beta Prec)$ means the Harmonic mean of $Precision$ and $Recall$. It is calculated as follows:
    
    \begin{equation}
        Harmonic(\alpha Rec, \beta Prec) = (\alpha +\beta )/(\frac{\alpha }{Rec} + \frac{\beta }{Prec})
    \end{equation}
    
    where \lllzy{$Prec$ represents the percentage of common words (between candidate responses and the references) in the candidate responses, and $Rec$ means the rate of common words in the references. Thus, $Prec$ reflects the accurate rate of the responses while $Rec$ indicate the loyalty to the references.} 
    To the end, we can calculate the LEPOR score based on Equation ~\ref{equ:lepor_1}. 
    The higher LEPOR score means the informativeness of the candidate is closer to that of the references.
    
\end{itemize}

For our study, we choose BLEU2 \cite{papineni2002bleu}, SLOR (syntactic log-odds ratio)\cite{kann2018sentence} and LEPOR \cite{han2012lepor} as our gold standard metrics in terms of adequacy, fluency and informativeness, respectively. 
Note that it is clear that this is an imperfect method for evaluating metrics as it assumes that the gold-standard metrics
represent the real users’ preferences. However, as pointed out by \citet{sakai2013metrics}, it is still useful to be able to quantify exactly
how often the metrics satisfy the basic properties, such as ``preference for a more
\emph{fluent} output'' or ``preference for a more \emph{informative} output'', without any explicit human judgments.

\subsubsection{Experiment settings} \label{sec:intuitivenesssettings}
\ 

In this section, we also consider SR metrics, including word overlap-based metrics (BLEU, METEOR and ROUGE), word embedding-based metrics (EA, SCS and BERTScore) and learning-based metrics (BERT-RUBER).\footnote{Here we do not present the results of BLEU3 and BLEU4 due to the limited space for the table. The variation between BLEU3 and BLEU4 is similar to the trend of BLEU2.} 
\lzy{Using the same models and settings as the SRST discriminative power experiment, we adopt 23 different systems (shown in Table \ref{tab:model_list}) to obtain the system runs. Thus, we have $(23 \times 22) / 2 = 253$ run pairs. As the MSDialog dataset contains 14,456 sessions and Wizardofwikipedia contains 22,197 sessions, we have $14,456 \times 253=3,657,368$ pairs of conversation responses in total for MSDialog and $22,197 \times 253=5,615,841$ pairs for Wizardofwikipedia.} 



\subsubsection{Experiment results} \label{sec:intuitivenessresults}
\ 

\begin{table}[]
    \centering
    \includegraphics[width=\textwidth]{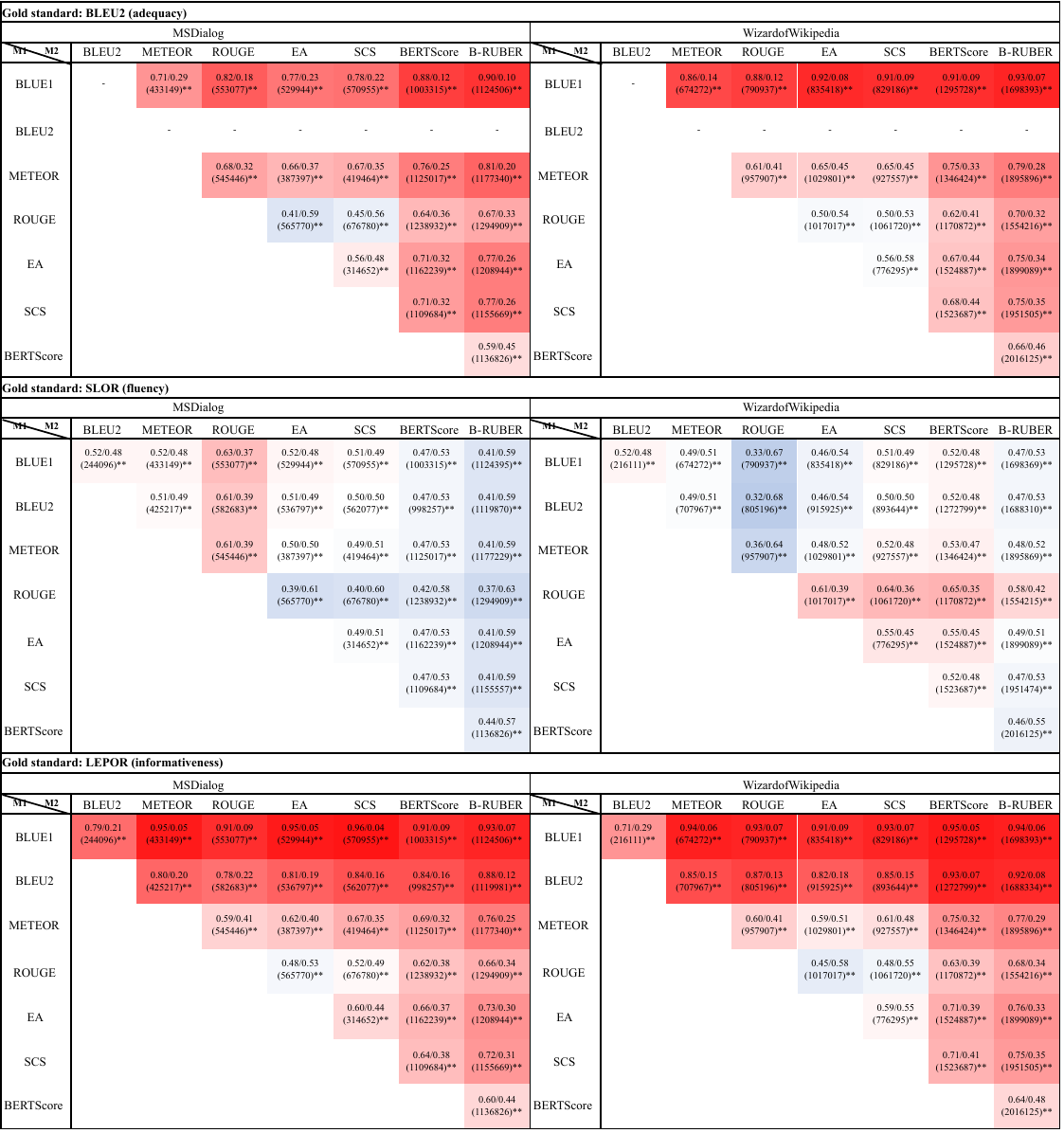}
    \caption{Concordance test results with the MSDialog and Wizardofwikipedia collection. \llzy{In each table cell, the upper left value indicates the intuitive score for the row metric, and the upper right value shows the intuitive score for the column metric. The value in the bottom bracket reveals the number of disagreements in terms of concordance with a given gold-standard metric} Two-tailed t-test is performed to detect any significant changes against each pair of metrics. * and ** represent significant value p < 0.05 and 0.01, respectively. The block color represents the concordance of direction compared with gold standard (red denotes the concordance score of metric M1 (the metric in the leftmost column) > M2 (the metric in topmost row) and blue represents the opposite, namely M1 < M2) while the brightness of the color indicates the (normalized) difference magnitude. \lllzy{Please note that BLEU2 is a gold standard and all the results which compare the concordance with BLEU2 could be 1.0/0.0 in adequacy test. Therefore, we do not show these results in the table.}}
    \label{tab:intuit}
\end{table}

\begin{table}[]
    \centering
    \includegraphics[width=\textwidth]{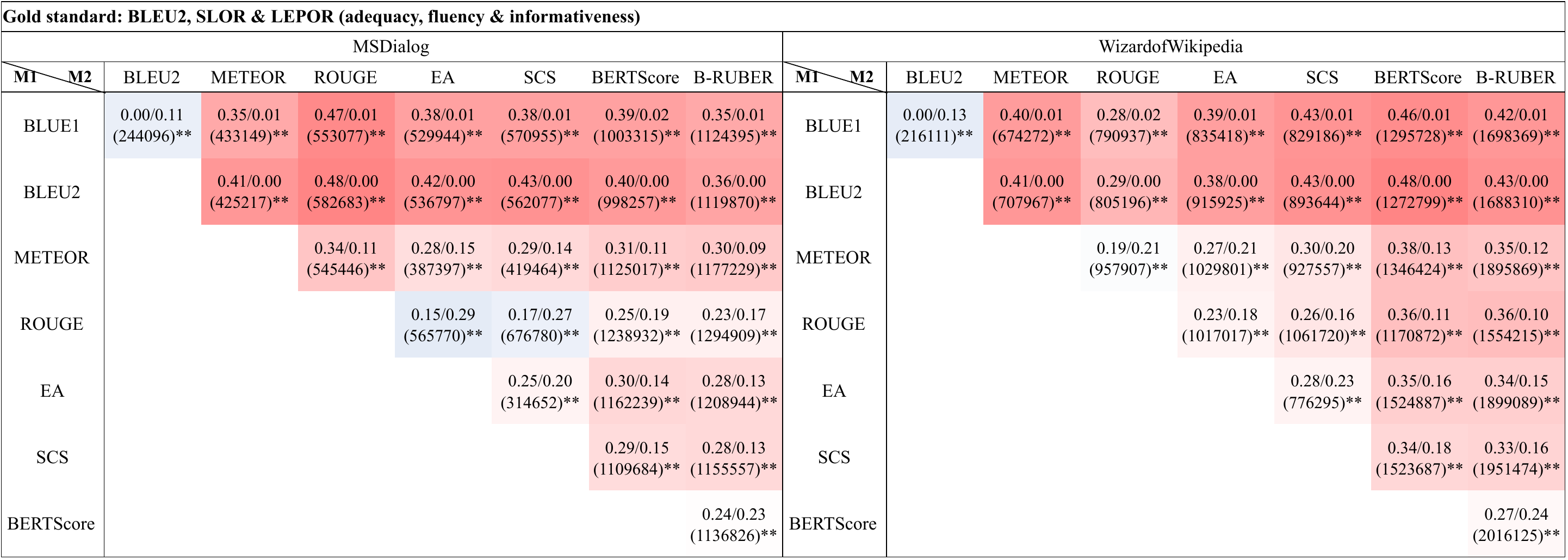}
    \caption{Concordance test results with the MSDialog and Wizardofwikipedia collection with all gold standard metrics. \llzy{In each table cell, the upper left value indicates the intuitive score for the row metric, and the upper right value shows the intuitive score for the column metric. The value in the bottom bracket reveals the number of disagreements in terms of concordance with a given gold-standard metric} Two-tailed t-test is performed to detect any significant changes against each pair of metrics. * and ** represent significant value p < 0.05 and 0.01, respectively. The block color represents the concordance of direction compared with gold standard (red denotes the concordance score of metric M1 (the metric in the leftmost column) > M2 (the metric in topmost row) and blue represents the opposite, namely M1 < M2) while the brightness of the color indicates the (normalized) difference magnitude}
    \label{tab:intuit_combine}
\end{table}

Table ~\ref{tab:intuit} shows the concordance test results of different metrics for the MSDialog and Wizardofwikipedia collections, respectively. These intuitiveness scores are computed using the preference agreement algorithm shown in Algorithm ~\ref{alg:ctalg}, 
%
which represent how the metrics favour the conversation responses with high adequacy (BLEU2), fluency (SLOR) or informativeness scores (LEPOR). 
For example, Table ~\ref{tab:intuit} shows the following information if we compare METEOR with ROUGE in terms of \emph{adequacy} in MSDialog corpus: (1) There are 545,446 disagreements between METEOR and ROUGE out of 3,657,368; (2) Of the 545,446 disagreements, METEOR is concordant with fluency 68\% of the time, while ROUGE is concordant with it 32\% of the time. That means given a pair of conversation responses for which METEOR and ROUGE disagree with each other, METEOR is more likely to agree with the adequacy evaluation than ROUGE. 

Let `M1 > M2' denote the relationship `M1 significantly outperforms M2 (p-value < 0.01) in terms of concordance with a given gold-standard metric'~\cite{zhou2013reliability}. It can be observed in Table ~\ref{tab:intuit} that:

(a) For \emph{adequacy}:
\begin{itemize}
    \item In MSDialog, BLEU1 > METEOR > EA > SCS > ROUGE > BERTScore > BERT-RUBER
    \item In WizardofWikipedia, BLEU1 > METEOR > SCS > EA > ROUGE > BERTScore > BERT-RUBER
\end{itemize}
\llzy{
The general trends of these two collections are very similar. Only the performance of EA and SCS vary in the two collections. It is not surprising to see that the BLEU1 metrics achieve the best performance because they share a similar algorithm and principles with the gold standard (i.e., BLEU2). Except for the BLEU1 metric, we find that METEOR is more intuitive than other metrics for adequacy.}

(b) For \emph{fluency}:
\begin{itemize}
    \item In MSDialog, BERT-RUBER > BERTScore > BLEU1 > BLEU2 = SCS > METEOR = EA > ROUGE
    \item In WizardofWikipedia, ROUGE > BERT-RUBER > EA >  METEOR > BLEU1 > SCS = BLEU2 > BERTScore

\end{itemize}
From the observations, we can find some similar trends across the two collections. Our findings are listed as follows:

\begin{itemize}
    
    \item \llzy{The leaning-based metric robustly outperforms other metrics across two collections. It can be observed that BERT-RUBER is ranked in the top 2 metrics for both MSDialog and WizardofWikipedia. This may be because the learning-based metric can capture more context information when the models are fine-tuned by the training datasets. Therefore, BERT-RUBER is more likely to agree with the fluency metric for evaluating conversation responses.}
    
    \item \llzy{Except for BERT-RUBER, the ranking of the other metrics varied considerably in different collections. For example, ROUGE performs the worst in MSDialog, but it performs the best in WizardofWikipedia. BERTScore outperforms most of metrics in MSDialog, while it gets the worst performance in WizardofWikipedia. From this perspective, both word overlap-based and embedding-based metrics are not able to robustly estimate the fluency in different collections.}
    
    \item \llzy{Except for the metric pairs including ROUGE}, the difference values in each metric pair are minor (less than 0.2), although these differences are significant. In other words, the performance of the metrics are very close in terms of fluency. 
    
    
\end{itemize}

(c) For \emph{informativeness}:
\begin{itemize}
    \item In MSDialog, BLEU1 > BLEU2 > METEOR > EA > ROUGE > SCS > BERTScore > BERT-RUBER
    \item In Wizardofwikipedia, BLEU1 > BLEU2 > METEOR > EA > SCS > ROUGE > BERTScore > BERT-RUBER
\end{itemize}
%
\llzy{The ranking of metrics in terms of informativeness are similar in the two collections. } The results show that the BLEU metric set (i.e., BLEU1 and BLEU2) achieves the best performance in both collections. This is not a surprise because BLEU only considers the different n-grams and may catch more informative words. Moreover, it can be also observed that word overlap-based metrics generally outperform the embedding-based metrics and learning-based metrics, which means that word overlapping may help catch more informativeness. 

\llzy{To take it a step further, we combine all the gold standard metrics together and evaluate the intuitiveness of the above metrics. Table ~\ref{tab:intuit_combine} shows the intuitiveness results. For the combination of all gold standard metrics,}

\begin{itemize}
    \item In MSDialog, BLEU2 > BLEU1 > METEOR > EA > SCS > ROUGE > BERTScore > BERT-RUBER
    \item In WizardofWikipedia, BLEU2 > BLEU1 > ROUGE > METEOR > EA > SCS > BERTScore > BERT-RUBER
\end{itemize}
\llzy{From the results, we can see the similar ranking of metrics in these two datasets. Generally, word overlap-based metrics achieve higher intuitiveness scores than word embedding-based and learning-based metrics in terms of all gold standard. It can be observed that although the learning-based metric BERT-RUBER could achieve good performance in terms of fluency, it still has poor performance in terms of overall gold standard.}

Overall, from Table ~\ref{tab:intuit} we can find every metric has its own strengths and weaknesses. For example, BLEU and METEOR can be more \textit{intuitive} in terms of adequacy and informativeness, but have poor performance for fluency. 

\lllzy{It is worthwhile noting that there could be some potential biases due to the gold standard metrics. Since the gold standards we choose in adequacy and intuitiveness are word overlap-based metrics, especially BLEU2 have a similar structure with other BLEU metrics (e.g., BLEU1), this setting may overestimate the intuitiveness of the overlap-based metrics. Moreover, except for fluency, the gold standard we used in the experiment are reference-based metrics, such as BLEU2 and LEPOR. The performance of these gold standards may also be influenced by the provided references. These are a possible limitation of our experiment. However, since few previous studies systematically proposed gold standard metrics in terms of these three dimensions, what we can do is to find the appropriate metrics from the related experiment results \cite{liu2016not, novikova2017we}. All the gold standard metrics we used have been proved to achieve the best performance in these previous studies \cite{liu2016not, novikova2017we}. Therefore, this is the best we can find for intuitiveness test with current support studies.}

%% file: content/4.4summaryofsingleturn.tex
Overall, we thoroughly discuss several representative metrics, including word overlap-based, word embedding-based metrics and a learning-based metric, in the single-turn conversational search scenario from different dimensions.

\begin{itemize}
    \item \textbf{Discriminative Power} describes how well the metrics can distinguish the difference in performance between different models. From our experiment, we find the presentation forms of responses may have a significant influence on the discriminative performance of the metrics. \llzy{For example, the word embedding-based metric BERTScore outperforms other metrics in SRST, but it has poor performance in MRST. }Specifically, based on the observations, we recommend METEOR if one is to evaluate conversational search responses in different scenarios, because METEOR can achieve relatively good and robust performance in both the single- and multi-response scenarios.
    
    \item \textbf{Predictive Power} measures the correlation between human preference and the metric prediction. If a metric has high predictive power, this metric is more likely to correctly predict human preference. Based on the experiments, we find {the learning-based metric BERT-RUBER can reach higher predictive scores, but its performance is not robust in different collections. Moreover, }METEOR also outperforms most of the other metrics in both collections. However, it is worth noting that the performance of most of the metrics (except for BERT-RUBER) are poor and only have minor improvement over a random selection (baseline). 
    
    \item \textbf{Intuitiveness} reflects how close the candidate metric is to the gold standard. We comprehensively test all SR metrics in terms of adequacy, fluency, and informativeness. Results show that every metric has its own strengths and weakness. BLEU and METEOR can perform  better for intuitiveness in terms of adequacy and informativeness, while they perform poorly in the fluency test. Further, we find that BLEU and METEOR have a balanced performance on these three aspects and may be a good choice if one is to evaluate the general performance in a conversational search scenario.
\end{itemize}

\lllzy{To sum up, METEOR may be, comparatively speaking, the best metric and a good choice in the single turn evaluation, when we consider all three perspectives.}

%% file: content/5multiturnevaluation.tex
In theory, we would follow our meta-evaluation framework and conduct our meta-evaluation on: discriminative power, predictive power, and intuitiveness. However, it is difficult to calculate the discriminative power of multi-turn dialogue systems because comparing the quality of dialogues from different multi-turn dialogue systems is not easy. For example, two different dialogue systems may generate quite different multi-turn responses even though the first utterances are the same. More complicated factors (e.g., the relevance of response in each turn, the patience of users) may influence the quality of the entire dialogue.
In terms of intuitiveness test,  there might be huge variance in the metric that looks at a single dimension (such as fluency) across multi-turns. Prior work has not examined how to calculate an individual dimension metric across multi-turn conversations. 

However, prior studies have been carried out to understand how user interactions in multi-turn conversations reflect the ultimate user satisfaction. This reflects the fidelity of the metric: whether the metrics are measuring what they aim to measure. Choi et al.~\cite{choi2019offline} propose a learning-based evaluation approach for multiple-turn dialogue systems. However, learning-based approaches may deeply rely on the training corpus and require user behavior as inputs, which may limit the application of this type of methods. 
Therefore, we carry out predictive power to investigate the fidelity of multi-turn metrics. 

\subsection{Methodology}


\lllzy{As described in \S\ref{sec:multipleturnmetric}, we assume that references for each round is already known. The variance of references in multi-turn conversations is not key concern in this paper. We only focus on evaluating the conversation when the question and references have been specific. Based on these specific questions and references, we adopt session-based metrics to predict the satisfaction of the conversations. }

Since we maintain absolute satisfaction labels of a multi-turn `session', rather than user preference data, we cannot use predictive power directly as described in algorithm ~\ref{alg:pdalg}. Instead, we use a concordance test (as algorithm ~\ref{alg:ctalg} in the intuitiveness test) and the gold standard satisfaction to calculate predictive power, following \cite{chen2017meta}.

The process of multi-turn meta-evaluation is: (1) Here we assume that metric scores can reflect the variation of response relevance in each turn. Based on our evaluation results (\S\ref{sec:summaryofsingleturn}), we select METEOR, which achieves the best performance in terms of three meta-evaluation aspects in ST, to estimate the `relevance' score of each response. Then the gain of each turn can be calculated by replacing relevance parameter $rel(q_i,r)$ with METEOR scores according to the Equation ~\ref{equ:gain}. (2) Regarding each 
single turn Q/A pair as a `special query' in a search session, we can estimate a score for the entire `session' based on a session-based metric framework. (3) Finally, a concordance test with real satisfaction scores are performed based on these estimated scores.



There are several common characteristics between session search and conversational search: (1) both of them are interactive searches; (2) they allow a user to perform multiple-rounds of interaction; (3) each round of interaction may have, to some extent, connections with other turns in a session. Therefore, session-based metrics might be appropriate alternative measures to evaluate conversational search.  For our study, instead of designing a new multiple-turn metric, we extend session-based evaluation approaches (shown in \S\ref{sec:multipleturnmetric}) to conversational search scenarios. The aim of this section is to answer the question: how do the session-based metrics perform in conversational search, and which ones are more suitable for evaluating conversational search systems? In order to answer this question, we conduct a concordance test by using these adapted search session-based metrics.

\subsection{Experiment Settings}

\lllzy{Here we only choose WizardofWikipedia to meta-evaluate the session-based metrics, since the MSDialog collection does not have users' satisfaction labels for the entire conversation. As described in \S\ref{sec:collection}, WizardofWikipedia has `Eval\_score' tags for each conversation, which are the satisfaction scores that users' provided. The satisfaction scores are integers which range from -1 to 5. We use these scores as the gold standard scores in the concordance test. Additionally, in order to compare the performance, we also adopt a random model, which randomly generates an integer score from -1 to 5 (the range is same as the original annotation in the collection) for each session.}

\subsection{Experiment Results}

\begin{table}[] 
\begin{tabular}{cc}
\toprule
\textbf{Model} & \textbf{Concordance} \\ \midrule
random & 0.4318 \\ \midrule
sCG & 0.5303** \\
sDCG & 0.5312** \\
snDCG/q & \lzyfinal{0.5350}** \\ \midrule
Decrease\_weight & 0.5315 \\
Increase\_weight & 0.5319** \\
Equal\_weight & 0.5329* \\
Middle\_high & 0.5286** \\
Middle\_low & 0.5339 \\ \midrule
Max & 0.5409** \\
Min & 0.4811** \\ \bottomrule
\end{tabular}
\caption{The concordance test between session-level satisfaction and the session-based metrics (* indicates two-tailed t-test statistical significance at p < 0.05 level with the random method, ** indicates t-test statistical significance at p < 0.01 level with the random method).}
\label{tab:multi}
\end{table}

Table ~\ref{tab:multi} shows the concordance test between satisfaction and the session-based metrics. The baseline is a random model. We perform a t-test for comparing the performance of each session-based metric with the baseline.
From this table, we can find most session-based metrics significantly outperform the random model. However, it can be observed that the difference between session-based metrics is very minor. Metric Max achieves the best performance among these metrics, which means the most relevant answer in the conversation may have a higher correlation with user satisfaction in the session-based evaluation framework. Interestingly, the Decrease\_weight and Middle\_low metrics do not show a significant difference with the baseline, while the Increase\_weight and Middle\_high rapidly improve over the baseline. To some extent, these preliminary results suggest that the relevance of the former responses in the conversation might have a weaker influence on the session-level satisfaction. 

It is worth noting that our basic assumption for multi-turn scenarios is conversational searches have some similar interaction with session-level searches, such as multiple-round interaction and effort cost during the search. However, it is obviously inadequate to regard conversational search as session search, since the interaction behavior in conversational searches might be more complex than that in session searches. From the experiment results, we can find all the session-based metrics weakly correlate with human annotations of satisfaction (all are less than 60\%), which also reflects the difference between conversational searches and session searches. Therefore, more specific behavior features of conversational search should be considered in the future metrics design.

%% file: content/6discussion.tex
In this paper, we have comprehensively measured the main representative conversational search metrics from different perspectives. Our results show that many metrics commonly used in previous work for evaluating dialogue systems are not good enough for conversational search. With no appropriate metrics for conversational search at the current stage, we provide some metric selection strategies, as well as suggestions based on our results. Generally, we find METEOR can achieve robust performance with various response forms in the search scenario. If an evaluation focuses on blended search environments (e.g., single responses or multiple responses), we recommend METEOR because it can robustly discriminate the difference between models and has relatively good correlation with human preferences across different collections. If evaluating the performance in the multi-round conversation, selecting the maximum metric value in the entire session might be a good choice for estimating the performance of models. 

Although more systematic and comprehensive than prior meta-evaluations of conversational search, there are still several limitations in our work: 
\begin{itemize}
    
    
    \item \lllzy{The systems/models (\S\ref{sec:systemruns}) applied in our experiment are extensive. In our experiment, we consider both retrieval and generative models, which covers most of search system application. However, it is still not comprehensive because several state-of-arts fine-grained approaches, such as a BERT-based system \cite{gulyaev2020goal}, are not considered in this paper. However, given that our selected systems cover a wide spectrum of differing performances and are of various types (i.e., retrieval and generative models), we believe that the incorporation of these new models would not bring significant changes to our meta-evaluation results.  }
    
    \item 
    \lllzy{There may be potential biases if the measures and the systems/models take similar approaches. For example, the measures might overestimate those systems when the metrics and systems/models share similar characteristics. To shed light on these biases, we rank all the systems/models (including retrieval based and generative models in Table ~\ref{tab:model_list}) and count the percentage of the top-1 ranked system that belongs to the type of retrieval or generative for each SR metric (shown in Table ~\ref{tab:metric}). We conduct such experiments for both MSDialog and WizardofWikipedia datasets. The results are shown in Table ~\ref{tab:metricbias}. In the table, for example, for BLUE1 on the MSDialog dataset, we find that 57\% of the top-1 ranked systems are retrieval based. From the results in Table ~\ref{tab:metricbias}, it can be observed that BERT-RUBER metric tends to rank generative models higher whereas ROGUE metric tends to favour retrieval models. These results provide insights that we need to be careful of when using those automatic metrics to evaluate systems. How to de-bias such effects is an interesting avenue for us to pursue in our future work. }
    
\begin{table}[h]
\begin{tabular}{|c|c|c|c|c|}
\hline
\multirow{2}{*}{} & \multicolumn{2}{c|}{MSDialog} & \multicolumn{2}{c|}{Wizard} \\ \cline{2-5} 
 & Retrieval & Generative & Retrieval & Generative \\ \hline
BLEU1 & 0.57 & 0.43 & 0.46 & 0.54 \\ \hline
BLEU2 & 0.55 & 0.45 & 0.43 & 0.57 \\ \hline
METEOR & 0.67 & 0.33 & 0.46 & 0.54 \\ \hline
ROUGE & \textbf{0.98} & 0.02 & \textbf{0.93} & 0.07 \\ \hline
EA & 0.67 & 0.33 & 0.55 & 0.45 \\ \hline
SCS & 0.59 & 0.41 & 0.42 & 0.58 \\ \hline
BERTScore & 0.23 & \textbf{0.77} & 0.40 & \textbf{0.60} \\ \hline
BERT-RUBER & 0.06 & \textbf{0.94} & 0.27 & \textbf{0.73} \\ \hline
\end{tabular}
\caption{The percentage of the top-1 model type in systems/models ranking for all SR metrics.}
\label{tab:metricbias}
\end{table}

    \item \lzy{In intuitiveness, it is worth noting that the gold standard is not completely equal to the human judgements. Although the gold standard metrics we chose perform the best and have relatively high correlation with human judgements in the corresponding work, they still have some limitations: (1) they are tested in different corpora and validated in different scenarios. The performance of gold standard may vary in a different corpus. (2) all the metrics used are not specific metrics for conversational search. Therefore, these metrics may not reflect all the characteristics of conversation search. To sum up, these gold standard metrics are not perfect, but we believe that they are well chosen for this meta-evaluation.}
    
    
    \item For the multiple-turn evaluation, we only directly use session-based search metrics to estimate the satisfaction of the whole conversation. Although conversational search and session search have some features in common, conversational search provides more complicated search interaction and behavior signals. Adopting session-based metrics directly may not be adequate for reflecting the real quality of the multiple-round interaction. 
    \item  In our study, we assume that only one ground truth response is available for each conversational context. Although this is common in most of the recent conversational search models, it is worth taking multiple ground truth responses into account when we evaluate the performance of conversational search models.

\end{itemize}

In this work, we do not propose new alternatives for evaluating conversational search. Considering the diversity of conversational search, it is really difficult for word-based metrics to cover all possible ground truth responses and evaluate the relevance only based on word matching. However, we believe that the metric based on context information and user behavior is a practicable direction. 

As for future work, we plan to define more common features of conversational search and better model typical conversational search scenarios. Beyond the existing word-based metrics, we plan to design sentence-level approaches for single-turn evaluation and try to apply context information for evaluation of multiple-turn conversational search. There are also four key research agendas that emerge from our results that can be studied in future work.

\lllzy{\textbf{Meta-evaluation with human annotations.} In this paper, we initially propose a meta-evaluation method for existing large-scale collections. We find that there are some limitations which are caused by the lack of human annotation. In future work, a deeper meta-evaluation could be performed by   collecting a human-labeled dataset that is specific for the meta-evaluation of different evaluation metrics. For example, we can ask the annotators to make pairwise preference annotations for response pairs and use them to evaluate the fidelity of evaluation metrics. We can also collect adequacy, fluency, and informativeness labels for responses and use them to investigate the intuitiveness of different metrics. }

\textbf{The impact of response presentation forms}. Our experiment shows the presentation forms of the response also significantly affect the performance of the evaluation metrics. As there is no uniform definition and interaction pattern for conversational search, the impact of response presentation forms is often ignored in previous work. Existing studies for evaluating dialogue system commonly treat the response as a single sentence \cite{liu2016not, lowe2016evaluation, lowe2017towards} or a paragraph \cite{cohen2018wikipassageqa}. In practice, the appearance of the response may vary widely in order to satisfy users' information needs. Here, we only consider the common appearance of the response used in conversational search. The results show that the distinguishing ability of the metrics may be changed when they are applied in different search scenarios (i.e., single-response or multiple responses). Therefore, considering the impact of response presentation forms is necessary when choosing appropriate conversational search metrics.

\textbf{Weak correlation between metrics and human judgements.} Based on the analysis of predictive power, our work shows that existing metrics do not correlate strongly with human judgements. This conclusion has been demonstrated in previous work for evaluating dialogue systems (e.g. \cite{liu2016not}). \lllzy{Therefore, the existing metrics may be unsuitable for estimating the human judgements in conversational search as well. How to reflect human preference is still a key concern in future metric design.} 


\textbf{Multiple turns evaluation.} To our best knowledge, there is no dedicated evaluation metrics for multiple-turn conversational search. As there are several common characteristics between multi-session search and conversational search, our study tries to measure the performance of several metrics which are commonly used in multi-session work. Results show conversational search does not follow the same `discounted rules' and all the evaluation frameworks perform poorly in the evaluation of multiple turns. It is worth noting that the Max strategy achieves the best performance, which means the users' satisfaction may be more likely to be affected by the best performance of QA pair in the conversation. However, we should note that there are some significant differences between multi-session search and conversational search. Directly adopting multi-session search metrics to evaluate conversational search in the future is perhaps not ideal. Although this is an alternative way to measure the multi-turn conversational search, future work should consider more specific features in conversational search evaluation. 

%% file: content/7conclusion.tex
In this article, we present a systematical meta-evaluation of conversational search metrics. A variety of conversational search metrics are comprehensively meta-evaluated in terms of \emph{reliability}, \emph{fidelity} and \emph{intuitiveness}. Moreover, we move beyond single-turn response settings and systematically test metrics across different scenarios, including single-turn and multi-turn search environments. The final goal of our work is to provide metric selection strategies in conversational search. According to our results, our main findings are:


\begin{itemize}
    \item In terms of reliability, METEOR \cite{banerjee2005meteor} and \llzy{BERTScore \cite{bert-score}} are capable of performing robustly in both SRST and MRST. When multiple responses are considered, RBP \cite{moffat2008rank} is able to achieve better discriminative power scores in capturing the difference of ranked responses.
    \item In terms of fidelity, we also observe a weak correlation between the existing metrics and human judgements as previous work \cite{liu2016not}. Results show that embedding-based metrics are generally less predictive than word overlap-based metrics across different test collections. \llzy{Besides, we also find the learning-based metric BERT-RUBER \cite{ghazarian2019better} can achieve higher correlation with human preference, but its performance may depend on the collections.}
    \item In terms of intuitiveness, we find that BLEU \cite{papineni2002bleu} and \llzy{METEOR \cite{banerjee2005meteor}} outperform other metrics in terms of adequacy and informativeness. \llzy{The learning-based metric BERT-RUBER \cite{ghazarian2019better} could be a good choice in capturing the fluency.}
    \item For the multiple-turn evaluation, we find the adopted session-based metrics are  moderately concordant with user satisfaction.
    
\end{itemize}

Finally, our work presents a general framework to conduct a meta-evaluation for conversational search using different test collections that can be adopted by others and expanded by the community. 